\documentclass[aps,preprint]{revtex4}

\usepackage{amsmath}
\usepackage{amssymb}
\usepackage[caption=false]{subfig} 
\usepackage{graphicx}
\usepackage[colorlinks=true,allcolors=blue]{hyperref}
\usepackage{soul}
\usepackage{todonotes}
\allowdisplaybreaks

\begin{document}
\preprint{MIT-CTP/6018}

\title{Critical dynamics of the superfluid phase transition
in Model F}

\author{Chandrodoy~Chattopadhyay$^1$, Robert~Maguire$^2$, 
Josh~Ott$^3$,  Thomas~Sch\"afer$^2$, and Vladimir~V.~Skokov$^{2}$}
\affiliation{$^1$ Theoretical Physics Division, Physical
Research Laboratory, Navrangpura, Ahmedabad 380009, India}
\affiliation{$^2$ Department of Physics and Astronomy, 
North Carolina State University, Raleigh, NC 27695}
\affiliation{$^3$ Center for Theoretical Physics -- a Leinweber Institute, 
Massachusetts Institute of Technology, Cambridge, MA 02139}

\begin{abstract}
We describe numerical simulations of the critical dynamics near
the superfluid phase transition. The calculations are based on 
an implementation of a stochastic hydrodynamic theory known as 
model F in the classification of Hohenberg and Halperin. This 
theory is expected to describe dynamic scaling near the lambda
transition in liquid $^4$He, Bose-Einstein condensation in ultracold 
atomic gases, and the superfluid transition in the unitary Fermi gas. 
Our simulation is based on a Metropolis algorithm previously applied
to the critical endpoint of the liquid-gas phase transition in ordinary
fluids. In the model E truncation of model F we obtain the expected
dynamical exponent $z\simeq 3/2$. We observe the emergence of a propagating
second sound mode at the phase transition. The second sound diffusivity
$D_s$ is consistent with the scaling relation $D_s\sim \xi^{x_\kappa}$,
where $\xi$ is the correlation length and $x_\kappa=1/2$.

\end{abstract}
\maketitle

%%%%%%%%%%%%%%%%%%%%%%%%%%%%%%%%%%%%%%%%%%%%%%%%%%%%%%%%%%%%%%%%%%%%%%
\section{Introduction}
\label{sec:intro}
%%%%%%%%%%%%%%%%%%%%%%%%%%%%%%%%%%%%%%%%%%%%%%%%%%%%%%%%%%%%%%%%%%%%%%

  The dilute Fermi gas at unitarity, a system of spin 1/2 fermions 
interacting via an $s$-wave interaction tuned to infinite scattering 
length, has emerged as an important model system for studying the 
transport properties of strongly correlated liquids \cite{Bloch:2008,
Schafer:2009dj,Adams:2012th,Schaefer:2014awa,Zwerger:2016}. In this 
work we are particularly interested in transport properties such as 
thermal relaxation and sound diffusivity in the vicinity of a second 
order phase transition. Our study builds on earlier work on critical 
transport in liquid helium \cite{Halperin:1976,Hohenberg:1977ym}, but 
it is motivated by more recent attempts to identify a critical point 
in the phase diagram of Quantum Chromodynamics (QCD) \cite{Bzdak:2019pkr}.

  The unitary Fermi gas has a second order phase transition between a 
superfluid phase at low temperature, and a normal phase at high 
temperature. This transition has been observed in both experiment
and in numerical simulations \cite{Ku:2012,Bulgac:2008zz}, and the 
available data is consistent with the prediction that the transition
is in the universality class of the three dimensional $O(2)$ model. 
Real time dynamics near the superfluid phase transition is expected 
to be governed by a hydrodynamic theory known as model F in the 
classification of Hohenberg and Halperin \cite{Hohenberg:1977ym,
Folk:2006ve}. Model F describes the coupled evolution of the order 
parameter and the diffusion of heat. It predicts that correlation 
functions exhibit dynamic scaling, and that the thermal conductivity 
and sound attenuation constant diverge near the critical temperature. 
Indications for this behavior were observed in experiments that study 
the linear response of the unitary Fermi gas to external perturbations
\cite{Yan:2024}.

  Superfluidity is also expected to occur in QCD in the regime of 
large baryon density. Two phases of matter that have been discussed
in the literature are superfluid neutrons at densities comparable 
to the density of ordinary nuclei, and quark superfluidity at even 
higher densities \cite{Dean:2003,Alford:2007xm}. The finite temperature
transitions associated with these phases are described by model F,
but it is not clear how the associated dynamics can be probed in the 
laboratory. The superfluid transition may have observable consequences
for the cooling of neutron stars \cite{Page:2011}. 

 A phase transition in QCD which is analogous to the superfluid transition
involves the restoration of chiral symmetry. This transition occurs between 
the hadronic phase at low temperatures and the quark-gluon plasma at high 
temperatures. The static universality class of this transition is
governed by an $O(4)$ model, and the dynamics is described by model G
\cite{Rajagopal:1992qz}. This hydrodynamic theory describes the coupling
of the order parameter with the conserved isospin currents in QCD. In 
particular, there is a Josephson relation that governs the evolution
of the phase of the order parameter. In QCD the chiral symmetry is 
explicitly broken by a quark mass. This corresponds to an external 
field coupled to the order parameter. It implies that the transition
is a smooth crossover, and that the correlation length remains finite.
It has nevertheless been argued that imprints of dynamic universality
can be observed in experiment \cite{Florio:2025zqv}.

 There has been a significant amount of work on dynamical scaling in model
F based on the epsilon expansion \cite{Halperin:1976,Dohm:1979,Dohm:1987,
Dohm:1991,Haussmann:1999,Dohm:2006}. These studies found that the dynamical 
exponent $z$ is exactly equal to $3/2$, and that the diffusivity of second
sound diverges as $D_s\sim \xi^{1/2}$, where $\xi$ is the correlation 
length. The goal of the present work is to study dynamical correlation 
functions in model F numerically. We have several motivations in mind. 
First, we want to verify the results of the epsilon expansion in a 
non-perturbative framework. Second, we want to construct a framework
that will allow us to compute non-universal quantities and make 
numerical predictions that can be confronted with cold gas 
experiments. 

  Not many numerical studies of critical dynamics near the superfluid
phase transition have appeared in the literature. We are only aware of 
the real-time statistical simulation of a spin model described in
\cite{Krech:1999}. Our work is based on recent advances in simulating
stochastic hydrodynamic theories, including model A (relaxational
dynamics) \cite{Schaefer:2022bfm}, model B (diffusion of a conserved
charge) \cite{Chattopadhyay:2023jfm}, model G (order parameter dynamics
in an anti-ferromagnet) \cite{Florio:2021jlx}, and model H (fluid
dynamics near a liquid-gas endpoint) \cite{Chattopadhyay:2024jlh}.
These simulations employ a Metropolis algorithm to ensure that the 
Langevin dynamics relaxes to the correct equilibrium distribution 
and satisfies fluctuation-dissipation relations. 

 This paper is structured as follows: We provide a brief introduction 
to model F of statistical dynamics in Sect.~\ref{sec:model-F}, describe
our numerical methods in Sect.~\ref{sec:num}, and present results on 
static and dynamic phenomena in Sect.~\ref{sec:res}. We give a brief
outlook in Sect.~\ref{sec:sum}. Numerical details regarding the 
parameters describing the static critical behavior are given in an 
appendix.
  
%%%%%%%%%%%%%%%%%%%%%%%%%%%%%%%%%%%%%%%%%%%%%%%%%%%%%%%%%%%%%%%%%%%%%%
\section{Model F}
\label{sec:model-F}
\subsection{Equations of Motion}
\label{sec:modF-EOM}
%%%%%%%%%%%%%%%%%%%%%%%%%%%%%%%%%%%%%%%%%%%%%%%%%%%%%%%%%%%%%%%%%%%%%%

 Model F describes the interaction between a complex order parameter 
$\phi(t,\vec{x})$ and a conserved density which we denote as $\psi
(t,\vec{x})$ \footnote{
Note that Hohenberg and Halperin use the notation $\psi$ for the
order parameter and denote the conserved density by $m$
\cite{Hohenberg:1977ym}.}.
As we will explain in more detail below, we can think of $\phi$ as
the wave function of the condensate, $\phi\sim\langle \psi_\uparrow
\psi_\downarrow\rangle$, and $\psi$ as proportional to the entropy 
per particle, $\psi\sim s/n$ \footnote{\label{q-def}
More precisely, $\psi$ is taken to be the variable that evolves
diffusively in linearized compressible fluid dynamics 
\cite{Kadanoff:1963}. This amounts to $d\psi=dq=Tn\,d(s/n)
=d\epsilon+(\epsilon+P)/n\, dn$, where $\epsilon$ is the energy 
density and $P$ is the pressure.}. 
Here, $\psi_\alpha$ with $\alpha=\ \uparrow,\downarrow$ is the 
Schr\"odinger field of a non-relativistic fermion. In a bosonic
superfluid $\phi\sim\langle\Phi\rangle$, where $\Phi$ is a 
non-relativistic scalar field. The equations of motion are given 
by \cite{Hohenberg:1977ym,Halperin:1976}
\begin{align}
 \frac{\partial \phi}{\partial t} &= - 2 \, \Gamma \,
 \frac{\delta {\cal H}}{\delta \phi^*} - i \, g_0 \, \phi \,
 \frac{\delta {\cal H}}{\delta \psi} + \theta\, ,  
 \label{evol_phi}  \\
 \frac{\partial \psi}{\partial t} &= \kappa \, \nabla^2 
 \frac{\delta {\cal H}}{\delta \psi} + 2 g_0 \, \mathrm{Im} \, 
 \left( \phi^* \, \frac{\delta {\cal H}}{\delta \phi^*} \right) 
 + \zeta. 
\label{evol_m}
\end{align} 
Here, $\Gamma$ is a complex relaxation rate, $g_0$ is the 
mode coupling parameter, $\kappa$ is the thermal conductivity, 
${\cal H}$ is the hydrodynamic Hamiltonian, and $\theta$ and
$\zeta$ are space-time dependent noise terms. We will take 
the Hamiltonian to be of the form
\begin{align}
 {\cal H}= \int d^3x \, 
  \left[ \frac{1}{2} \, |\nabla \phi|^2 
     + \frac{1}{2} \, m^2 \, |\phi|^2 
     + \frac{\lambda}{4} \, |\phi|^4 
     + \frac{1}{2 \, C_0} \, \psi^2 
     + \gamma_0 \, \psi \, |\phi|^2 
     - \mathrm{Re}(H \, \phi^*) - H_\psi \, \psi \right],
\label{Hamiltonian}
\end{align}
where $m$ is the inverse bare correlation length, $\lambda$
is the self coupling of the order parameter, $C_0$ is the bare
specific heat, $\gamma_0$ is the coupling of the conserved
density to the order parameter, and $H$ as well as $H_\psi$ 
are external fields. The noise terms are zero on average.
The correlation function of the noise is given by 
\begin{align}
\langle \theta(t, \vec{x}) \, \theta^*(t', \vec{x}') \rangle 
  &= 4 \, T  \, \mathrm{Re} \,(\Gamma) \, 
  \delta\left( \vec{x} - \vec{x}' \right) \, \delta(t - t'), 
\label{noise_theta} \\
 \langle \zeta(t, \vec{x}) \, \zeta(t', \vec{x}') \rangle 
 &= - 2 \, \kappa \, T \, \nabla^2 
 \delta \left( \vec{x} - \vec{x}' \right) \, \delta(t - t'), 
 \label{noise_zeta} \\
\langle \theta \theta \rangle 
  &= \langle \theta^* \theta^* \rangle 
   = \langle \theta \zeta \rangle 
   = \langle \theta^* \zeta \rangle = 0,
\label{noise_theta_cross_terms}
\end{align} 
where $T$ is the temperature. The equation of motion for $\psi$ 
can be written in a manifestly conserving form. To see this, we 
note that
\begin{align}
\phi^* \, \frac{\delta {\cal H}}{\delta \phi^*} 
 = - \frac{1}{2} \, \phi^* \, \nabla^2 \phi + f(|\phi|) 
 = - \frac{1}{2} \, \vec{\nabla} \cdot 
   \left( \phi^* \, \vec{\nabla} \phi \right)  
    + \frac{1}{2} |\nabla \phi|^2 + f(|\phi|), 
\end{align}
where $f(|\phi|)$ is a real function,
\begin{align}
   f(|\phi|) = \frac{1}{2} \, m^2 |\phi|^2 
      + \frac{\lambda}{2} \, |\phi|^4 
      + \gamma_0 \, \psi \, |\phi|^2\, . 
\end{align}
We conclude that 
\begin{align}
 \mathrm{Im} \, \left( \phi^* \, 
   \frac{\delta {\cal H}}{\delta \phi^*} \right) 
   = - \frac{1}{2} \, \vec{\nabla} \cdot \, 
      \mathrm{Im} \left( \phi^* \, \vec{\nabla} \phi \right), 
\end{align}
so that Eq. (\ref{evol_m}) can be written as 
\begin{align}
\label{psi-cons}
\frac{\partial \psi}{\partial t} + \vec{\nabla} \cdot \vec{\jmath} 
      = 0, \quad\quad\quad
\vec{\jmath} = - \kappa \, \vec{\nabla} \, 
  \frac{\delta {\cal H}}{\delta \psi} 
+ g_0 \, \mathrm{Im} \, \left( \phi^* \, \vec{\nabla} \phi \right) 
+ \vec{\xi},
\end{align}
where the noise $\vec{\xi}$ is also delta-correlated,
\begin{align}
\langle \xi_i(t, \vec{x}) \, \xi_j(t', \vec{x}') \rangle 
 = 2 \, \kappa \, T \, \delta_{ij} \, \delta(\vec{x} - \vec{x}') 
 \delta(t - t').
\end{align}
If the external fields are included then the density $\psi$
is no longer conserved, $\partial_t\psi + \vec{\nabla} \cdot
\vec{\jmath} = - g_0 \, \mathrm{Im} (\phi^* \, H)$.    

%%%%%%%%%%%%%%%%%%%%%%%%%%%%%%%%%%%%%%%%%%%%%%%%%%%%%%%%%%%%%%%%%%%%%
\subsection{Equations of motion in $O(2)$ notation}
\label{sec:modF-EOM-O2}
%%%%%%%%%%%%%%%%%%%%%%%%%%%%%%%%%%%%%%%%%%%%%%%%%%%%%%%%%%%%%%%%%%%%%

 For the numerical implementation, we will represent the order 
parameter in terms of real-valued fields $(\phi_1, \phi_2)$
defined by 
\begin{align}
\label{phi-12}
\phi_1 = \frac{1}{2} \, \left( \phi + \phi^* \right), \quad
\phi_2 = \frac{i}{2} \, \left( \phi^* - \phi \right).
\end{align}
This implies, for example, that the rhs.~of Eq.~(\ref{evol_phi})
is 
\begin{align}
    \frac{\delta {\cal H}}{\delta \phi^*} 
    = \frac{1}{2} \, \left( \frac{\delta {\cal H}}{\delta \phi_1} 
    + i \, \frac{\delta {\cal H}}{\delta \phi_2} \right).
\end{align}
We will also decompose the noise $\theta = \theta_1 + i \theta_2$ 
and the external field $H = H_1 + i H_2$ into $O(2)$ components. 
Eqs.~(\ref{noise_theta}, \ref{noise_theta_cross_terms}) imply that
the noise correlators of $\theta_1$ and $\theta_2$ are equal and 
governed by the real part of $\Gamma$. The off-diagonal correlators
vanish and we have
\begin{align}
    \langle \theta_1(t, \vec{x}) \, \theta_1(t', \vec{x}') \rangle 
 &= \langle \theta_2(t, \vec{x}) \, \theta_2(t', \vec{x}') \rangle 
  = 2 \, \Gamma_1 \, T \, \delta 
   \left( \vec{x} - \vec{x}' \right) \, \delta(t - t'). \\
   \langle \theta_1(t, \vec{x}) \, \theta_2(t', \vec{x}') \rangle 
 &= 0.
\end{align}
The real and imaginary parts of Eq.~(\ref{evol_phi}) determine 
the equations of motion for $(\phi_1, \phi_2)$,
\begin{align}
    \frac{\partial \phi_1}{\partial t} &= - \Gamma_1 \, 
    \frac{\delta {\cal H}}{\delta \phi_1} + \Gamma_2 \, 
    \frac{\delta {\cal H}}{\delta \phi_2} \, + g_0 \, \phi_2 \, 
    \frac{\delta {\cal H}}{\delta \psi} + \theta_1, \\
    \frac{\partial \phi_2}{\partial t} &= - \Gamma_1 \, 
    \frac{\delta {\cal H}}{\delta \phi_2} - \Gamma_2 \, 
    \frac{\delta {\cal H}}{\delta \phi_1} \, - g_0 \, \phi_1 \, 
    \frac{\delta {\cal H}}{\delta \psi} + \theta_2.
\end{align}
Here, $\Gamma_1= \mathrm{Re}(\Gamma)$ is the real part of
$\Gamma$, and $\Gamma_2=\mathrm{Im}(\Gamma)$ is the imaginary 
part.

%%%%%%%%%%%%%%%%%%%%%%%%%%%%%%%%%%%%%%%%%%%%%%%%%%%%%%%%%%%%%%%%%%%
\subsection{Physical interpretations and theoretical expectations: 
Static behavior}
\label{sec:ModelF-stat}
%%%%%%%%%%%%%%%%%%%%%%%%%%%%%%%%%%%%%%%%%%%%%%%%%%%%%%%%%%%%%%%%%%%

  In thermodynamic equilibrium the fields are sampled from
the Gibbs distribution $p[\phi_a,\psi]\sim\exp(-{\cal H}[\phi_a,
\psi]/T)$. The partition function is 
\begin{align}
   Z(H_a,H_\psi,T) = \int D\phi_a\, D\psi \;
     \exp\left(-{\cal H}[\phi_a,\psi]/T\right).
\end{align}
Note that the dependence on $\psi$ is Gaussian, and
for static observables the conserved density can be integrated out.
The parameters $C_0,\gamma_0$ merely shift the value of $\lambda$.
The resulting Hamiltonian ${\cal H}[\phi_a]$ is that of the 
$O(2)$ model, also known as the $xy$ model,
\begin{align}
 {\cal H}_{xy}= \int d^3x \,  \left[ 
   \frac{1}{2} \, \left(\nabla \phi_a\right)\left(\nabla\phi_a\right)
     + \frac{1}{2} \, m^2 \, \phi_a\phi_a
     + \frac{\lambda'}{4} \, \left(\phi_a\phi_a\right)^2
     - H_a \, \phi_a \right],
\label{H-xy}
\end{align}
where $\lambda'=\lambda-2C_0\gamma_0^2$ is the shifted valued 
of $\lambda$. The critical exponents of the $O(2)$ model are 
well known \cite{Zinn-Justin:2002ecy},
\begin{align}
\begin{array}{c|ccccccc}
     & \alpha & \beta   & \gamma & \delta & \nu  & \eta & \omega\\ \hline
Z_2  &  0.110 & \;0.326 & \; 1.237 & \; 4.790 & \; 0.630 & \; 0.036 & \; 0.830\\
O(2) & -0.015 &   0.349 &    1.318 &    4.779 &    0.672 &    0.038 & \; 0.794
\end{array}
\label{crit-exp}
\end{align}
where for comparison we also show the exponents of the $Z_2$
(Ising) model. Here, $\alpha$ is the heat capacity exponent, 
$C\sim\tau^{-\alpha}$, $\beta$ governs the order parameter, 
$\Sigma\sim(-\tau)^\beta$, and $\gamma$ is the susceptibility exponent,
$\chi\sim\tau^{-\gamma}$. We have defined $\tau=m^2-m_c^2$, where 
$m_c$ is the critical value of $m$. At the critical point, $\delta$ 
is the order parameter exponent, $\Sigma\sim h^\delta$ and $\eta$
is the correlation function exponent, $C(x)\sim x^{2-d-\eta}$.
Finally, $\nu$ controls the correlation length, $\xi\sim
|\tau|^{-\nu}$. The leading correction-to-scaling exponent 
$\omega$ for the $O(2)$ model was determined in 
\cite{Chester:2019ifh}.

 We observe that difference between the $xy$ and Ising exponents
is very small. The main distinction is that in the $xy$ model 
$\alpha$ is small and negative, whereas in the Ising model 
$\alpha$ is small and positive. This means that in the $xy$
model the heat capacity has a non-analyticity, but not a 
divergence at the critical point. This is the origin of the 
term ``$\lambda$ transition''. 

 The order parameter $\Sigma$ is defined as 
\begin{align}
\label{M-def}
\Sigma = \lim_{H_1\to 0} \lim_{V\to \infty} 
  \left\langle M_1  \right\rangle\, , \quad
M_a =  \frac{1}{V}\int d^3x \, \phi_a(x) \, ,
\end{align}
where $V$ is the volume of the system. The order parameter
defines a magnetic equation of state
\begin{align}
 \Sigma = h^{1/\delta}f_G(y)+f_{\it reg}(H_1,m^2)\, , 
\end{align}
where $f_G(y)$ is the universal, singular, part and $f_{\it 
reg}(H,m^2)$ is the regular part of the equation of state. 
Here
\begin{align}
 y= \tau h^{-1/(\beta\delta)}\, , \quad
 \tau = \frac{m^2-m_c^2}{\tau_0}\,, \quad
 h = \frac{H_1}{H_1^0}\, , 
\end{align}
where $H^0_1$ and $\tau_0$ are normalization factors, defined
by $f_G(0)=1$. In a system at finite volume there is a 
generalized equation \cite{Karsch:2023pga}
\begin{align}
 \langle M_1\rangle  = h^{1/\delta}\left[f_G(y,y_L)
   +h^{\omega\nu_c}f_G^{(1)}(y,y_L)\right]\, , 
\end{align}
where $y_L=(L/L_0)h^{\nu_c}$, $L_0$ is a non-universal 
constant, $\nu_c=\nu/(\beta\delta)$, $\omega$ is the 
first sub-leading exponent, and we have neglected regular
terms.

 We can define a susceptibility matrix
\begin{align}
\chi_{ab} = 
  \frac{\partial^2\log(Z)}{\partial H_a\partial H_b} \, .
\end{align}     
In the restored phase $O(2)$ invariance implies that $\chi_{ab}
=\delta_{ab}\chi$ as $H_a\to 0$. If we approach $\tau=0$ from 
above $\chi$ diverges as $\tau^{-\gamma}$. In the broken phase
\begin{align}
\chi_{ab} = \left(\delta_{ab}-\hat{n}_a\hat{n}_b\right)\chi_T
 + \hat{n}_a\hat{n}_b\chi_L\, ,
\end{align}
where $\hat{n}_a=H_a/H$ with $H^2=(H_a)^2$. The transverse 
susceptibility diverges as $H\to 0$, $\chi_T=\Sigma/H$ for any 
value of $\tau<0$. The zero field longitudinal susceptibility 
has the same critical scaling as $\chi$, $\chi_L\sim 
(-\tau)^{-\gamma}$. In the broken phase the long-wavelength
dynamics is governed by an effective theory that only 
contains the phase $\varphi$ of the order parameter
\cite{Halperin:1976,Hasenfratz:1989pk}
\begin{align}
{\cal H}_{\varphi} = \int d^3x\, \left[
\frac{f^2}{2}(\nabla\varphi)^2 
  + \frac{f^2m_{\varphi}^2}{2} \varphi^2 
\right]\, ,
\label{GB-EFT}
\end{align}
where $f$ and $m_\varphi$ are the Goldstone boson decay constant
and mass. In Eq.~(\ref{GB-EFT}) we have omitted terms that contain
additional gradients or powers of $m_\varphi$. The Goldstone
boson mass is given by the Gell-Mann-Oakes-Renner relation
$f^2m^2_\varphi=H\Sigma$. In the theory of superfluidity 
$v_s=\nabla\varphi$ (in units $m_B=\hbar=1$) is called the 
superfluid velocity, and $f^2=\rho_s$ is the superfluid
mass density. The superfluid mass current is $\jmath=\rho_s
v_s$. 

  In the symmetric phase we can define the correlation length
in terms of the correlation function of the order parameter,
$\langle \phi^a(0)\phi^b(x)\rangle=\delta^{ab}G_\phi(x)$. At
large $x$ we expect $G_\phi(x)\sim\exp(-x/\xi)$. Near the 
critical point $\xi\sim\tau^{-\nu}$. In the broken phase the
existence of a Goldstone boson implies that the naively
defined correlation length is infinite as $H\to 0$. It is 
conventional to define a correlation length $\xi_T\sim 1/f^2$
(in $d=3$) \cite{Hohenberg:1977ym}, and one can show that 
this correlation length scales as $\xi_T\sim (-\tau)^{-\nu}$
\cite{Fisher:1973}.

%%%%%%%%%%%%%%%%%%%%%%%%%%%%%%%%%%%%%%%%%%%%%%%%%%%%%%%%%%%%%%%%%%%
\subsection{Dynamic behavior}
\label{sec:ModelF-dyn}
%%%%%%%%%%%%%%%%%%%%%%%%%%%%%%%%%%%%%%%%%%%%%%%%%%%%%%%%%%%%%%%%%%%

 We begin by discussing the physical meaning of the terms
in the model F equations of motion, and the truncations
that are involved in reducing the full set of hydrodynamic 
equations of a superfluid to model F. Consider the equations
of motion without mode coupling terms, 
\begin{align}
 \frac{\partial \phi}{\partial t} &= - 2 \, 
  \left(\Gamma_1+i\Gamma_2\right) \,
 \frac{\delta {\cal H}}{\delta \phi^*} + \theta\, ,  
 \label{evol_phi_0}  \\
 \frac{\partial \psi}{\partial t} &= \kappa \, \nabla^2 
 \frac{\delta {\cal H}}{\delta \psi} + \zeta. 
\label{evol_m_0}
\end{align} 
For $\Gamma_1\neq0$, $\Gamma_2=0$ Eq.~(\ref{evol_phi_0}) is a 
relaxation equation for a non-conserved order parameter (model 
A). This theory has a dynamic exponent close to $z=2$. In 
the broken phase there is a Goldstone boson which corresponds
to the phase of $\phi$. This mode is purely diffusive, not only 
at $T_c$, but also in the broken phase. For $\Gamma_2\neq 0$
Eq.~(\ref{evol_phi_0}) is a stochastic Gross-Pitaevskii (GP)
equation \cite{Stoof:2001,Gardiner:2002}. For a weakly 
interacting condensate of bosons of mass $m_B=2m_A$, where 
$m_A$ is the mass of the fermionic atom, we can match the 
gradient term in the Hamiltonian to the Schr\"odinger
equation by setting $\Gamma_2=\hbar/(2m_B)$. The potential 
terms can be viewed as a chemical potential for $\phi$. The 
stochastic GP equation near a critical point was studied in
\cite{Tauber:2014}, and it was shown that the critical dynamics
corresponds to model A.

 Eq.~(\ref{evol_m_0}) is a diffusion equation for the 
conserved density $\psi$. Once the dynamical equation 
for $\psi$ is included we can no longer integrate out 
the conserved density, and the dynamical scaling behavior 
near the critical point is modified from model A universality
to model C. In practice, this makes very little difference. 
For $\alpha<0$, as is the case in the $O(2)$ model, the 
dynamic exponent $z$ of model C is equal to that of 
model A. Even if $\alpha$ is positive (as in the Ising
model), the dynamic exponent changes from $z=2+c\eta$
(with a small coefficient $c$) to $z=2+\alpha$, with a
small exponent $\alpha$.

 The most important ingredient in model F is the 
presence of mode couplings,
\begin{align}
\frac{\partial \phi}{\partial t} = 
      \left\{{\cal H},\phi\right\}\, ,
   \quad\quad
\frac{\partial \psi}{\partial t} = 
    \left\{{\cal H},\psi\right\}\, .
\end{align}
The fundamental Poisson bracket is taken to be 
\begin{align}
\label{Poisson-fund}
 \left\{ \phi,\psi\right\} = ig_0\phi\, , 
\end{align}
which, together with $\{\phi^*,\psi\}=-ig_0\phi^*$, leads to
the mode-coupling terms in Eq.~(\ref{evol_phi},\ref{evol_m}).
The physical significance of $g_0$ becomes clear if we 
assume that Eq.~(\ref{Poisson-fund}) is derived from the 
quantum mechanical commutator between the particle density
$n$ and the field operator, $[n(x),\phi(x')]=2\phi\,\delta(x-x')$,
where we have assumed that $\phi\sim\psi_\uparrow
\psi_\downarrow$ carries particle number 2. Then we
can write
\begin{align}
\label{Josephson}
\frac{\partial \phi}{\partial t} = 
      \left\{ {\cal H},\phi\right\} = 
  \int \frac{\delta {\cal H}}{\delta n}\,  
     \left\{n,\phi\right\} = 2i\mu \phi\, , 
\end{align}
where $\mu$ is the chemical potential. Eq.~(\ref{Josephson})
is the Josephson equation, and taking the fundamental 
Poisson bracket to be $\{n,\phi\}=2i\phi$ implies $g_0=
-2(\partial\psi/\partial n)=2[(s/n)T+\mu]$, where we have used
thermodynamic relations for the variable $d\psi=nT\,d(s/n)$.

 The mode coupling equation for $\psi$ can be written as a 
 conservation equation $\partial_t\psi+\nabla\jmath=0$ with 
 $\vec\jmath=g_0\,{\rm Im}(\phi^*\vec\nabla\phi)$, see 
Eq.~(\ref{psi-cons}). In the regime of validity of the effective
theory Eq.~(\ref{GB-EFT}) we have $\vec\jmath=g_0\rho_s\vec{v}_s$. 
This result is counterintuitive, because, based on the definition 
of the density $\psi$, the current $\vec\jmath$ must be proportional 
to the entropy current. However, in superfluid hydrodynamics entropy 
is carried by the normal fluid, whereas $\rho_sv_s$ is the superfluid
mass current. The resolution of this puzzle is that in model F we 
view ordinary sound as a high frequency mode, and therefore ignore 
the dynamics of the momentum density. If the total momentum $\vec{\pi}
=\rho_n \vec{v}_n+\rho_s\vec{v}_s$ is frozen, then fluctuations
of $\vec\jmath$ are indeed proportional to $\rho_n\vec{v}_n$.

 In the broken phase, the mutual mode coupling between the order 
parameter and the conserved density leads to a propagating mode 
which mixes the phase of the order parameter with $\psi$. In the 
regime of validity of the Goldstone boson EFT this mode has 
the dispersion relation
\begin{align}
\omega =c_s \sqrt{k^2+m_\varphi^2}\, , \quad\quad
 c_s^2=\frac{g^2_0\rho_s}{C_0}\, . 
\end{align}
This is a second sound mode, an oscillation of the entropy density 
in which the superfluid moves relative to the normal fluid. As 
expected, the speed of sound goes to zero at the critical point. 
Hohenberg and Halperin observed that the existence of a propagating 
mode determines the dynamical critical exponent. Using $\rho_s\sim
\xi_T^{-1}$ and demanding that the dispersion relation is 
consistent with $\omega\sim \xi^{-z}$ for $k\xi_T\sim {\it const}$
we obtain $z=3/2$ (for $d=3$). Here, we have used the fact that 
$\alpha<0$, so that there is no divergence in $C_0$. Taking the 
dissipative terms into account, the second sound mode acquires
a width $w=c_sk-(i/2)\, D_sk^2$, where we have set $m_\varphi=0$.
Compatibility with dynamic scaling implies that $D_s\sim \xi_T^{1/2}$,
and since $C_0$ remains finite, we have $\kappa\sim D_S\sim \xi_T^{1/2}$.
These scaling relations have been checked in the context of the 
epsilon expansion \cite{Halperin:1976}.

  Finally, we note that there is a truncation of model F that is
known as model E. In this theory we set the coupling $\gamma_0$
as well as the imaginary part of $\Gamma$ to zero. Calculations 
within the context of the renormalization group indicate that this 
simplification does not change the universal critical dynamics
\cite{Halperin:1976,Folk:2006ve}. In the case of setting $\gamma_0
=0$ this is related to the fact that in the $O(2)$ model the specific 
heat does not diverge, and a purely Gaussian Hamiltonian for $\psi$
is sufficient. The reason that $\Gamma_2$ can be set to zero is 
that, as explained above, the critical dynamics of the stochastic
GP equations is the same as that of model A/C.

%%%%%%%%%%%%%%%%%%%%%%%%%%%%%%%%%%%%%%%%%%%%%%%%%%%%%%%%%%%%%%%%%%%
\section{Numerical implementation}
\label{sec:num}
\subsection{Lattice formulation}
\label{sec:lat}
%%%%%%%%%%%%%%%%%%%%%%%%%%%%%%%%%%%%%%%%%%%%%%%%%%%%%%%%%%%%%%%%%%%

 We discretize the fields $\phi_a$ and $\psi$ on a spatial lattice
with lattice spacing $a$, and adopt a set of units where $a=1$. We 
employ periodic boundary conditions on a cubic lattice with sides of 
length $L=Na$. The discretized Hamiltonian is 
\begin{align}
{\cal H}  =& \sum_{\vec{x}} \Bigg[ \frac{1}{2}  
    \sum_{\mu=1}^d   (\phi_a(\vec{x}+\hat{\mu}) - \phi_a(\vec{x}) )^2  
    +  \frac{1}{2} m^2 \phi_a^2(\vec{x}) 
    +   \frac{1}{4} \lambda  \left(\phi_a^2(\vec{x})\right)^2  
    - H_a \phi_a (\vec{x})  \nonumber \\ 
\label{H-lat}
 & \quad\quad \mbox{}+ \frac{1}{2C_0}\,\psi(\vec{x})^2
     + \gamma_0\,\psi(\vec{x})\phi_a^2(\vec{x})
     - H_\psi \psi(\vec{x})\Bigg]\, , 
\end{align}
where we have suppressed the time argument of the fields and the 
sum over the index $a=1,2$. The sum over $\vec{x}$ is a sum over 
integer vectors $\vec{n}$ with $\vec{x}=a\vec{n}$. We also define 
a unit vector $\hat \mu$ in the direction $\mu\in \{ 1,\ldots, d \}$, 
where $d$ is the number of spatial dimensions. In the present work 
we will only consider $d=3$. Following our earlier work on model H
\cite{Chattopadhyay:2024jlh}, as well as previous work on model G
\cite{Florio:2021jlx}, we will separate the time update into a 
diffusive step and an ideal step. The diffusive step is performed
using a Metropolis algorithm which is designed to lead to the 
correct equilibrium distribution of the fields $\phi_a$ and $\psi$,
and to respect fluctuation-dissipation relations. 

%%%%%%%%%%%%%%%%%%%%%%%%%%%%%%%%%%%%%%%%%%%%%%%%%%%%%%%%%%%%%%%%%%%
\subsection{Metropolis update}
\label{sec:MC}
%%%%%%%%%%%%%%%%%%%%%%%%%%%%%%%%%%%%%%%%%%%%%%%%%%%%%%%%%%%%%%%%%%%

We first consider the dissipative/stochastic evolution of $\phi_a$
\begin{align}
\label{mod-A}
    \frac{\partial \phi_{a}}{\partial t} = 
     - \Gamma_1 \, \frac{\delta {\cal H}}{\delta \phi_a} + \theta_a.
\end{align}
The required Metropolis scheme is identical to that in Model A
\cite{Schaefer:2022bfm}. For every site $\vec{x}$ in the lattice we 
choose a trial update
\begin{align}
\label{phi-MC}
    \phi_a^{\mathrm{new}}(\vec{x}) = \phi_a^{\mathrm{old}}(\vec{x}) 
    + \sqrt{2 \,  T \, \Gamma_1 \, \Delta t}\, \Theta,
\end{align}
where $\Theta$ is a random number drawn from a normal distribution
with zero mean and unit variance. We compute the change in Hamiltonian 
due to this trial update $\Delta {\cal H}$ for the field $\phi_a$
and accept with probability $P = {\it max}(1,e^{-\Delta{\cal H}/T})$. 
If the update is accepted,
\begin{align}
    \phi_a(t + \Delta t, \vec{x}) = \phi_a^{\mathrm{new}}(\vec{x}),
\end{align}
else the field is kept unchanged. For the conserved density $\psi$, the
diffusive/stochastic part of the evolution is
\begin{align}
\label{mod-B}
 \frac{\partial \psi}{\partial t} 
    + \vec{\nabla} \cdot \vec{\jmath}_{\mathrm{diss}} = 0\, , 
    \quad\quad
    \vec{\jmath}_{\mathrm{diss}} = 
       - \kappa\,\vec\nabla \, \frac{\delta {\cal H}}{\delta \psi} 
       + \vec{\xi}.
\end{align}
Here the Metropolis step is constructed in analogy with the 
model B update described in \cite{Chattopadhyay:2023jfm}. The basic
idea is to integrate the evolution equation over the cell centered 
at $\vec{x}$
\begin{align}
\partial_t \psi(t,\vec{x})  = 
  - \sum_{\mu = 1}^{\mu = 3} \, \left( q_\mu^{+} - q_\mu^{-} \right), 
\end{align}
where $q_{\mu}^{\pm}$ are fluxes through forward and backward faces 
of the cell. A trial update involves a pair of cells centered at 
$(\vec{x}, \vec{x} + \hat{\mu})$, 
\begin{align}
\label{psi-MC}
\psi_{\mathrm{new}}(\vec{x}) = 
  \psi_{\mathrm{old}}(\vec{x}) + q_\mu\, , \quad 
\psi_{\mathrm{new}}(\vec{x} + \hat{\mu}) = 
  \psi_{\mathrm{old}}(\vec{x} + \hat{\mu}) - q_\mu\, , \quad
q_\mu = \sqrt{2 \, \kappa \, T \, \Delta t} \, \lambda,
\end{align}
where $\lambda$ is drawn from a distribution with zero mean and unit
variance. Again, we compute the change $\Delta {\cal H}$ due to the 
proposed update, and accept with probability $P={\it max}(1,e^{-\Delta
{\cal H}/T})$. As explained in our earlier work, the Metropolis updates 
have the property that the mean updates $\langle[\phi_a(t+\Delta t,x)-
\phi_a(t,x)]\rangle$ and $\langle[\psi(t+\Delta t,x)-\psi(t,x)]\rangle$ 
realize the relaxation and diffusive terms in Eqs.~(\ref{mod-A},\ref{mod-B}).
Furthermore, the variance of the updates reproduces the variance of 
the stochastic forces.

%%%%%%%%%%%%%%%%%%%%%%%%%%%%%%%%%%%%%%%%%%%%%%%%%%%%%%%%%%%%%%%%%%%%%%%%%%%%%%
\subsection{Non-dissipative evolution}
\label{sec:ideal}
%%%%%%%%%%%%%%%%%%%%%%%%%%%%%%%%%%%%%%%%%%%%%%%%%%%%%%%%%%%%%%%%%%%%%%%%%%%%%%

 The ideal, non-dissipative, part of the evolution equations is 
\begin{align}
\dot{\phi}_1 &= 
 \Gamma_2 \, \Big[  - \nabla^2\phi_2 + m^2 \, \phi_2 
   + \lambda \, |\phi|^2 \, \phi_2 + 2 \, \gamma_0 \, \psi \, \phi_2 
   -  H_2 \Big] \nonumber \\
   & \quad\quad\quad \mbox{}
   + g_0 \, \phi_2 \, \left( \frac{\psi}{C_0} + \gamma_0 \, |\phi|^2 
   - H_\psi  \right), \label{evol_phi1_explicit}\\
\dot{\phi}_2 &= 
- \Gamma_2 \, \Big[  - \nabla^2\phi_1 + m^2 \, \phi_1 
  + \lambda \, |\phi|^2 \, \phi_1 + 2 \, \gamma_0 \, \psi \, \phi_1 
  - H_1 \Big] \nonumber \\
  & \quad\quad\quad \mbox{}
  - g_0 \, \phi_1 \, \left( \frac{\psi}{C_0} + \gamma_0 \, |\phi|^2 
  - H_\psi  \right), 
  \label{evol_phi2_explicit}\\
\dot{\psi} &= 
 - g_0 \, \left( \phi_1 \, \nabla^2 \phi_2 
     - \phi_2 \, \nabla^2 \phi_1 \right) 
 -  g_0 \, \left( \phi_1 \, H_2 - \phi_2 \, H_1 \right)\, . 
\label{evol_psi_explicit}
\end{align} 
Eq.~(\ref{evol_psi_explicit}) explicitly conserves
the integral of $\psi$ if $H_a=0$. Furthermore, the 
coupled set of equations conserves the hydrodynamic
Hamiltonian ${\cal H}$,
\begin{align}
\frac{d{\cal H}}{dt} &= 
  \int d^3x \, \Big[ - \dot{\phi}_1 \, \nabla^2 \phi_1 
                     - \dot{\phi}_2 \, \nabla^2 \phi_2 
    + \left( \dot{\phi}_1 \, \phi_1 
    + \dot{\phi}_2 \, \phi_2 \right) \, 
    \left( m^2 + \lambda \, |\phi|^2 
    + 2 \, \gamma_0 \, \psi \right) 
    \nonumber \\
   & + \frac{1}{C_0} \, \psi \, \dot{\psi} 
     + \gamma_0 \, |\phi|^2 \, \dot{\psi}  
     -  H_1 \, \dot{\phi}_1 - H_2 \, \dot{\phi}_2 
     - H_\psi \, \dot{\psi}
    \Big] = 0 \, , 
\label{dH_dt} 
\end{align}
which follows simply from substituting 
Eqs.~(\ref{evol_phi1_explicit}-\ref{evol_psi_explicit}) into
Eq.~(\ref{dH_dt}). More fundamentally, the conservation of 
${\cal H}$ is due to the fact that the equations of motion
are generated by Poisson brackets, so that $\dot{\cal H}=
\{{\cal H},{\cal H}\}=0$. Our goal is to find a discretization
scheme for the equations of motion in which conservation 
of $\psi$ and ${\cal H}$ in the ideal step is exact even if 
the lattice spacing $a$ is finite. 

 In this context we observe that in deriving Eq.~(\ref{dH_dt})
we did not use any identities other than integration by parts,
$\int d^3x\, (\nabla \phi_a)^2 = - \int d^3x \, \phi_a \, 
\nabla^2 \, \phi_a$. Motivated by this observation we define
forward and backward derivatives in the spatial direction 
$\mu=1,2,3$ as 
\begin{align}
\nabla^{R}_{\mu}\phi(\vec{x}) = \frac{1}{a}
 \left[\phi(\vec{x}+\hat{\mu}a)-\phi(\vec{x}) \right],
 \hspace{0.5cm}
\nabla^{L}_{\mu}\phi(\vec{x}) = \frac{1}{a}
 \left[\phi(\vec{x})-\phi(\vec{x}-\hat{\mu}a) \right].
\end{align}
The Laplacian is defined as $\nabla^2=\nabla^L_\mu\nabla^R_\mu
= \nabla^R_\mu\nabla^L_\mu$, where summation over $\mu$ is 
implied. Note that this lattice derivative satisfies 
integration
by parts 
\begin{align}
    \sum_{\vec{x}} \nabla^R_\mu\phi(\vec{x})
     \nabla^R_\mu\phi(\vec{x})
 =  \sum_{\vec{x}} \nabla^L_\mu\phi(\vec{x})
     \nabla^L_\mu\phi(\vec{x})
 = -\sum_{\vec{x}} \phi(\vec{x})
     \nabla^2\phi(\vec{x}).
\end{align}
We then define the right hand side of the equations of 
motion (\ref{evol_phi1_explicit}-\ref{evol_psi_explicit})
using the lattice derivative $\nabla^2$. We integrate 
the equations over a time step $\Delta t$ using the 
strongly stable third order Runge-Kutta scheme of 
Shu and Osher \cite{Shu:1988}. Finally, we note that 
there is a lattice version of the non-dissipative part
of the conserved current $\jmath$, 
\begin{align}
\label{cur-lat}
\jmath_\mu(\vec{x}) = g_0 \left( 
    \phi_1(\vec{x})\nabla^R_\mu\phi_2(\vec{x})
  - \phi_2(\vec{x})\nabla^R_\mu\phi_1(\vec{x})\right)\, ,
\end{align}
which has the property that $\dot\psi(t,\vec{x})=-\nabla^L_\mu
\,\jmath_\mu(t,\vec{x})$. Note that we can also define a 
current $\jmath^L_\mu$ using the backwards derivative 
$\nabla_\mu^L$ so that $\dot\psi(t,\vec{x})=-\nabla^R_\mu
\,\jmath^L_\mu(t,\vec{x})$.

%%%%%%%%%%%%%%%%%%%%%%%%%%%%%%%%%%%%%%%%%%%%%%%%%%%%%%%%%%%%%%%%%%%%%%%%%%%%%%
\begin{figure}[t]
\centering
\includegraphics[width=0.45\linewidth]{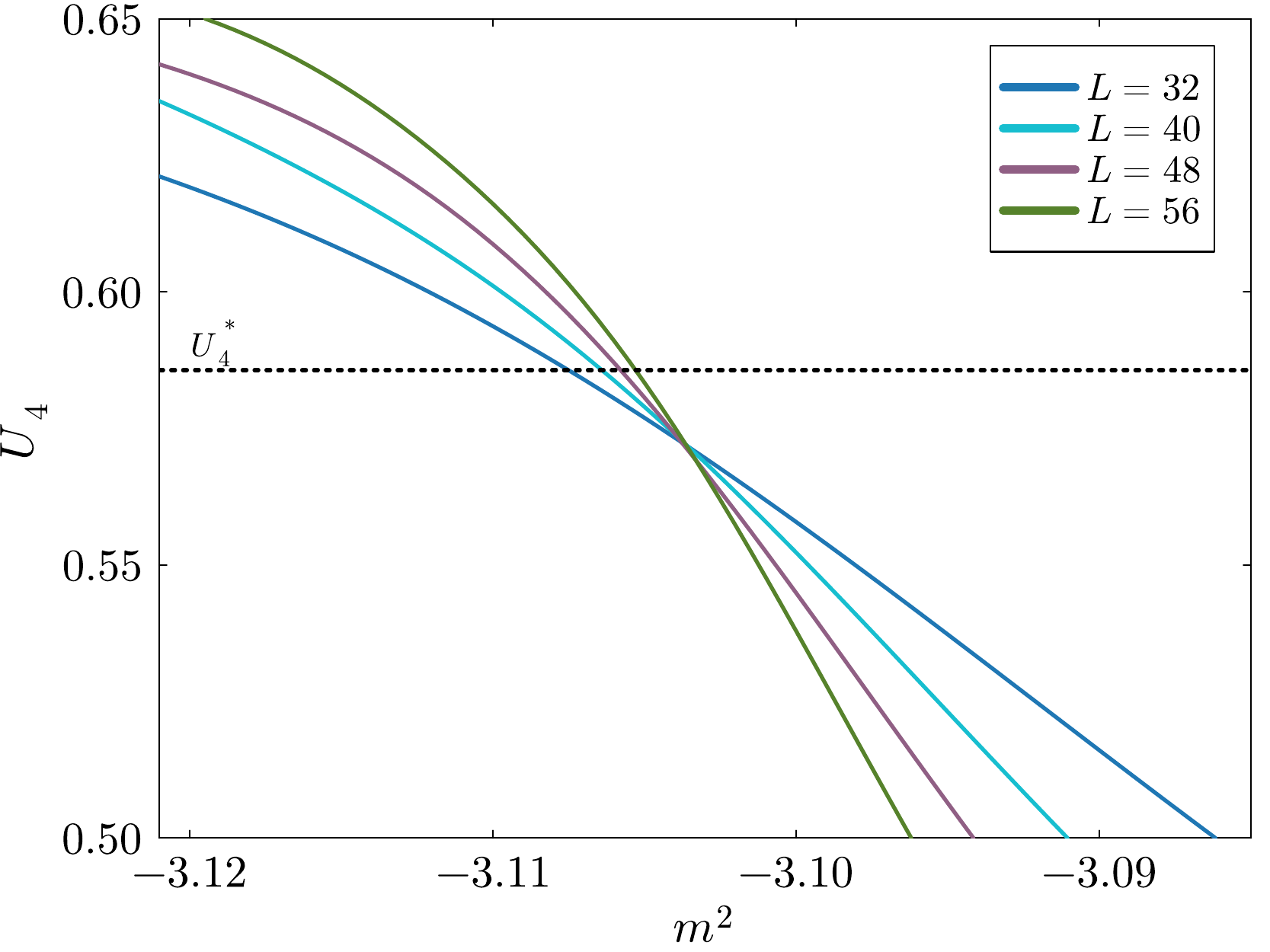}
\includegraphics[width=0.45\linewidth]{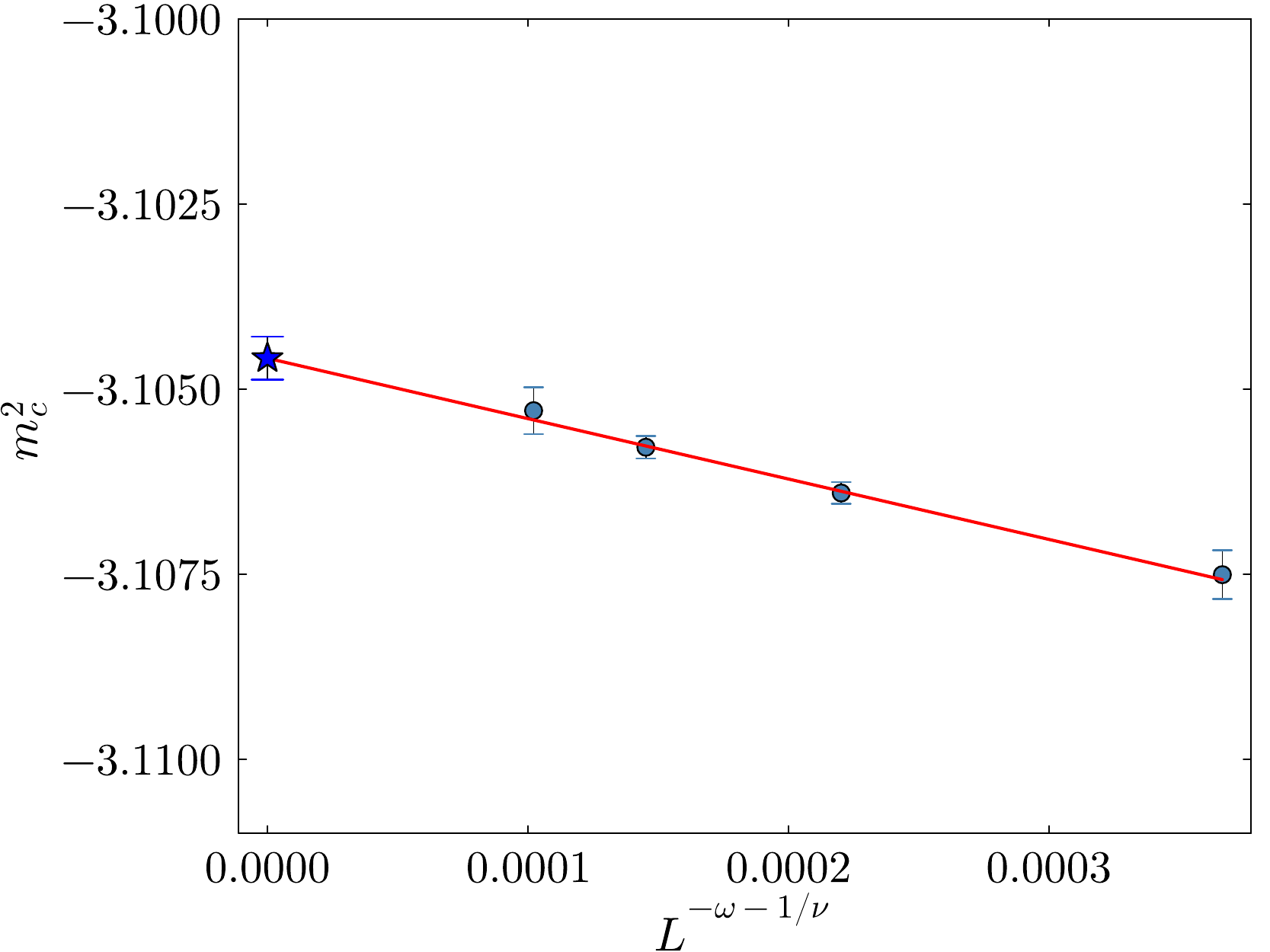}
\caption{Left panel: The binder cumulant $U_4$ as a function of 
temperature $m^2$ for several values of the lattice size $L$. The 
universal value, $U_4^*$, is shown by the dashed black line.  Right panel: The 
value of $m^2$ where $U_4$ crosses the critical value as a function
of the scaled inverse lattice size $L$. The star indicates the 
infinite volume extrapolation, $m_c^2 =  -3.1046(3)$.}
\label{fig:binder}
\end{figure}
%%%%%%%%%%%%%%%%%%%%%%%%%%%%%%%%%%%%%%%%%%%%%%%%%%%%%%%%%%%%%%%%%%%%%%%%%%%%%%

%%%%%%%%%%%%%%%%%%%%%%%%%%%%%%%%%%%%%%%%%%%%%%%%%%%%%%%%%%%%%%%%%%%%%%%%%%%%%%
\section{Numerical Results}
\label{sec:res}
\subsection{Static behavior}
\label{sec:res-stat}
%%%%%%%%%%%%%%%%%%%%%%%%%%%%%%%%%%%%%%%%%%%%%%%%%%%%%%%%%%%%%%%%%%%%%%%%%%%%%%

 We first study the static behavior of model F. We have to do this 
in order to locate the critical value of $m^2$, and to verify that
we obtain the correct static scaling behavior of the $O(2)$ model. 
In the process we also obtain the non-universal parameters $H_0$ 
and $\tau_0$ (which we will define below) that will allow us to map
the $O(2)$ model onto a realistic equation of state. As explained
in Sect.~\ref{sec:ModelF-stat} the static behavior of model F is 
the same as that of model A/C. Indeed, since the critical exponent 
$\alpha$ is negative model C reduces to model A, and we can set the 
coupling $\gamma_0$ between the order parameter and the conserved 
density to zero. We use the dynamical algorithm described in
Sect.~\ref{sec:num}. Note that this algorithm describes critical 
slowing down. As a result, it does not provide an efficient method
for simulating the static behavior. Much more accurate results 
can be obtained with algorithms that avoid critical slowing down, 
such as a cluster algorithm.

%%%%%%%%%%%%%%%%%%%%%%%%%%%%%%%%%%%%%%%%%%%%%%%%%%%%%%%%%%%%%%%%%%%%%%%%%%%%%%
\begin{figure}[t]
\centering
\includegraphics[width=0.5\linewidth]{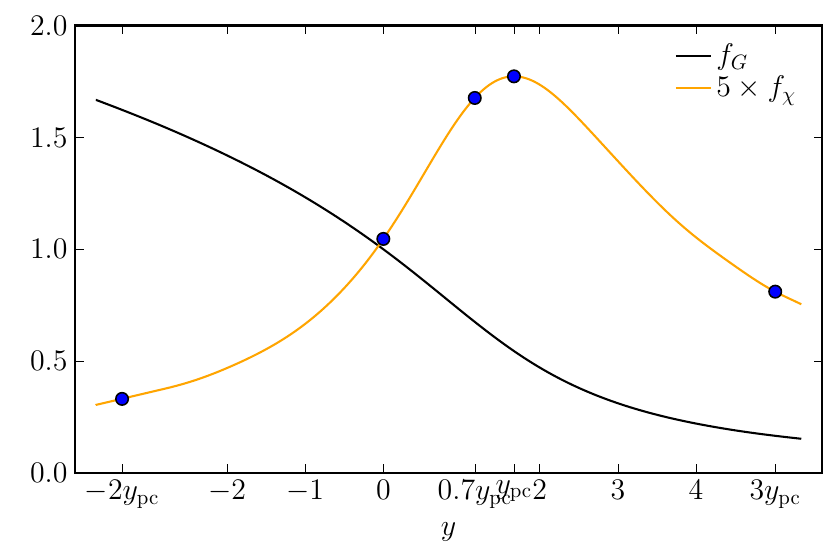}
\caption{The magnetic equation of state and susceptibility as a function 
of the scaling variable $y$. The dots depict the values of $y$ at which 
we have simulated the dynamics of the system. Here, $y_{\rm pc} \approx 
1.671$ is the position  of the susceptibility peak.
}
\label{fig:fgfchi}
\end{figure}
%%%%%%%%%%%%%%%%%%%%%%%%%%%%%%%%%%%%%%%%%%%%%%%%%%%%%%%%%%%%%%%%%%%%%%%%%%%%%%

 Our methods for determining $m_c^2$ and analyzing static criticality
closely follow our earlier work on model A \cite{Schaefer:2022bfm}.
In that case we considered the lattice Hamiltonian Eq.~(\ref{H-lat})
for a single real field. Using Binder cumulants we obtained $m_c^2=
-2.285$ for $\lambda=4$. The Binder cumulant is defined by 
\begin{align}
 U_4 = 1 - \frac{\langle (M_a^2)^2\rangle}{3 \langle M_a^2\rangle^2}\, ,
\end{align}
where the magnetization $M_a$ is defined in Eq.~(\ref{M-def}).
In the left panel of Fig.~\ref{fig:binder} we show results for the 
Binder cumulant using model F simulations with the $O(2)$ Hamiltonian 
for $\lambda=4$ and different values of $m^2$.
In practice we performed 
a number of simulations on coarse lattices to determine the approximate 
crossing point, and then used reweighting to obtain results on 
larger lattices. 
The reweighting is performed using the reference $m_{\rm ref}^2 = - 3.098$.
The horizontal line indicates the universal critical
value of the Binder cumulant, $U_4^*=0.585$ \cite{Campostrini:2000iw}.
In the right panel we show the value of $m^2$ at which the Binder 
cumulant $U_4(m^2,L)$ intersects the line $U_4 = U_4^*$. Finite size
scaling predicts $m^2(L) = m^2_c + C L^{-1/\nu-\omega}$. We observe
a linear relation between $m^2(L)$ and $L^{-1/\nu-\omega}$ and 
extrapolate to the critical $m^2$ in the infinite volume limit, 
$m_c^2=-3.1046(3)$, where the stated uncertainty is purely statistical. 

In App.~\ref{sec:App-H0} we discuss the extraction of the parameters
$H_0$, $\tau_0$, and $L_0$. We find
\begin{align}
 y = \frac{m^2-m_c^2}{\tau_0}\, 
 \left(\frac{H}{H_0}\right)^{-1/(\beta\delta)}\, , \quad
 \tau_0= 4.8\, , \quad
 H_0=7.97\, , 
\end{align}
and $y_L=(L/L_0)h^{\nu_c}$ with $L_0=0.567$. We can use these
quantities to identify specific points on the universal equation
of state at which we will perform dynamical simulations. In 
Fig.~\ref{fig:fgfchi} we show the magnetic equation of state 
and the universal susceptibility curve, taken from a 
parametrization of the $O(2)$ equation of state described in 
\cite{Karsch:2023pga}. We overlay values of $y$, listed in 
Table \ref{tab:compset}, at which we have performed 
simulations.

%%%%%%%%%%%%%%%%%%%%%%%%%%%%%%%%%%%%%%%%%%%%%%%%%%%%%%%%%%%%%%%%%%%%%%%%%%%%%%
\begin{table}[h]
\centering
\begin{tabular}{c|c|c|c|c|c|}
$y $ & $-2 y_{\rm pc}$  & $0$ & $0.7 y_{\rm pc}$ & $y_{\rm pc}$ 
                                     & $3 y_{\rm pc}$\; \\ \hline 
$m^2$\; & $-3.3$\; & $-3.1046$\; & $-3.04$\; & $-3$\; & $-2.8$\;  
\end{tabular}
\caption{Simulation parameters for dynamical studies performed 
in this work. All simulations are performed at $H = 0.005$, except 
in the case $y=0$, where we consider several values of $H$.}
\label{tab:compset}
\end{table}
%%%%%%%%%%%%%%%%%%%%%%%%%%%%%%%%%%%%%%%%%%%%%%%%%%%%%%%%%%%%%%%%%%%%%%%%%%%%%%

%%%%%%%%%%%%%%%%%%%%%%%%%%%%%%%%%%%%%%%%%%%%%%%%%%%%%%%%%%%%%%%%%%%%%%%%%%%%%%
\begin{figure}[t]
\centering
\includegraphics[width=0.49\linewidth]{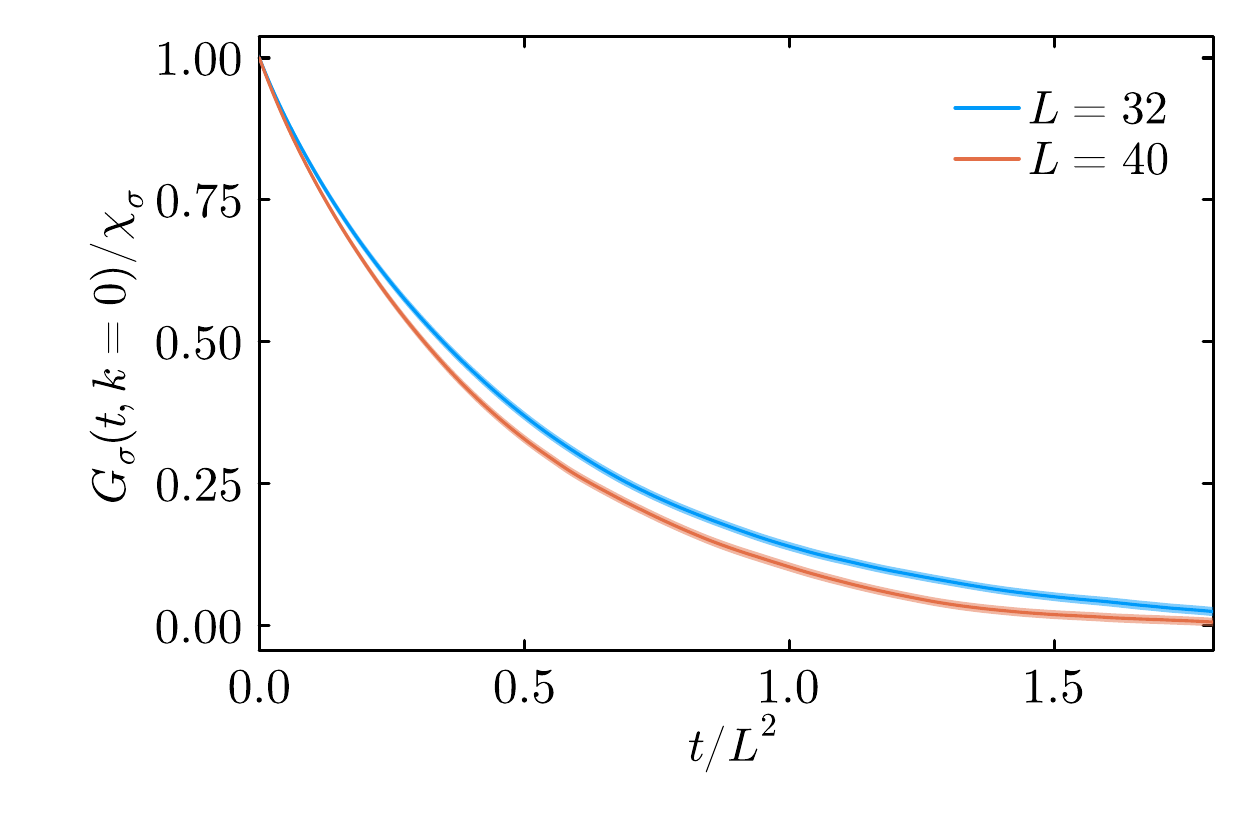}
\includegraphics[width=0.49\linewidth]{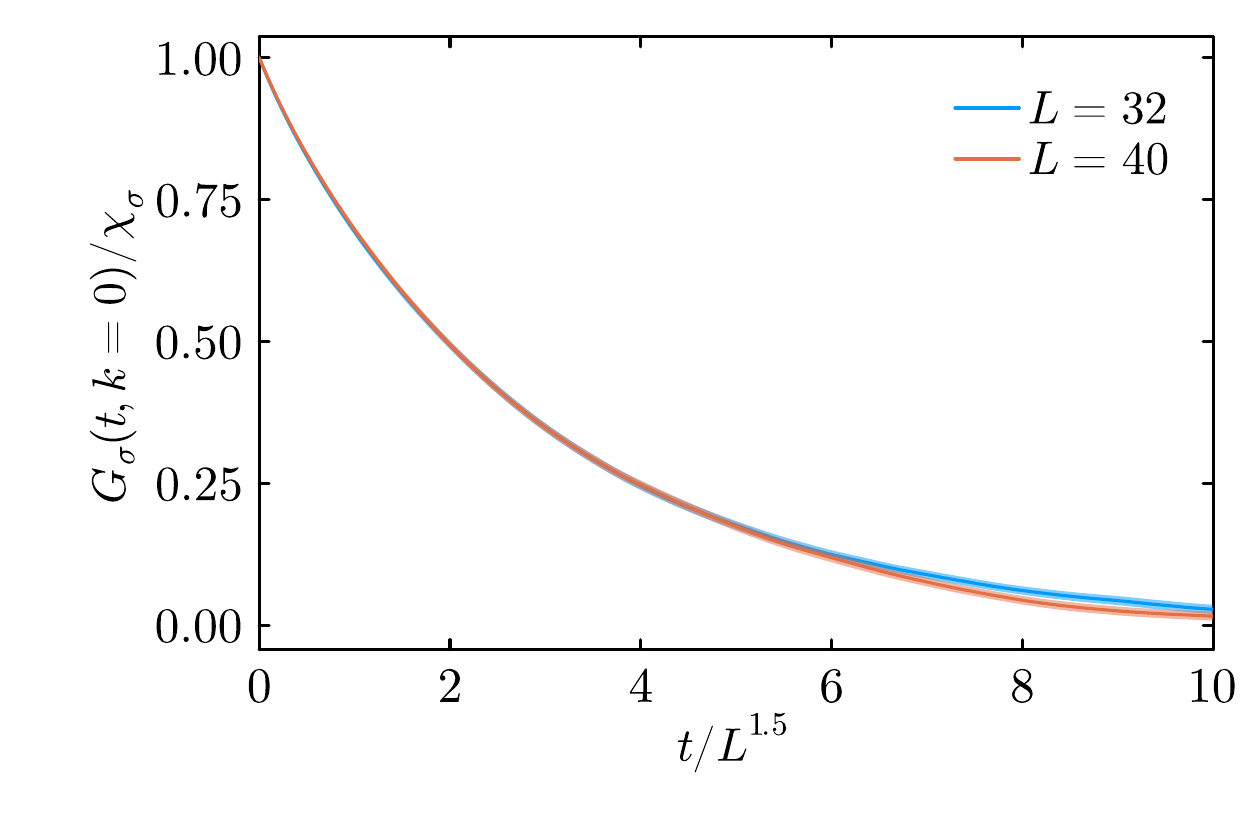}
\caption{Order parameter correlation function $G_\sigma(t,0)$
at the critical point $m^2=m_c^2$ and $H=0$ for two different values 
of the lattice size, $L=32$ and $L=40$. The correlation function
is plotted as a function of the scaling variable $t/L^z$ for two
different values of the dynamic exponent, $z=2$ (left panel) and
$z=3/2$ (right panel). The best fit value of $z$, based on 
the scaling behavior at $G_\sigma(t)/\chi_\sigma = 0.3$ is 
$z = 1.51 \pm 0.14$.}
\label{fig:Z_extr_phi}
\end{figure}
%%%%%%%%%%%%%%%%%%%%%%%%%%%%%%%%%%%%%%%%%%%%%%%%%%%%%%%%%%%%%%%%%%%%%%%%%%%%%%

%%%%%%%%%%%%%%%%%%%%%%%%%%%%%%%%%%%%%%%%%%%%%%%%%%%%%%%%%%%%%%%%%%%%%%%%%%%%%%
\subsection{Dynamics}
\label{sec:res-dyn}
%%%%%%%%%%%%%%%%%%%%%%%%%%%%%%%%%%%%%%%%%%%%%%%%%%%%%%%%%%%%%%%%%%%%%%%%%%%%%%

  In this Section we describe dynamical simulations based on the 
algorithm described in Sect.~\ref{sec:num}. We have set $\gamma_0=
\Gamma_2=0$ and focus on model E dynamics, as explained in 
Sect.~\ref{sec:ModelF-dyn}. We can absorb $C_0$ into the normalization
of $\psi$ and set $C_0=1$. We define a reference temperature $T_0$ so 
that the dimensions of the fields are $[\phi_a]=(T_0/a)^{1/2}$ and
$[\psi]=(T_0/a^3)^{1/2}$. We use the order parameter relaxation 
rate to define the unit of time and set $\Gamma \equiv \Gamma_1
\equiv 1$. We then have two physical parameters, the mode coupling 
$g_0$ and the thermal conductivity $\kappa$. In the following we 
show results for $(T/T_0)=g_0=\kappa=1$.

%%%%%%%%%%%%%%%%%%%%%%%%%%%%%%%%%%%%%%%%%%%%%%%%%%%%%%%%%%%%%%%%%%%%%%%%%%%%%%
\begin{figure}[t]
\centering
\includegraphics[width=0.49\linewidth]{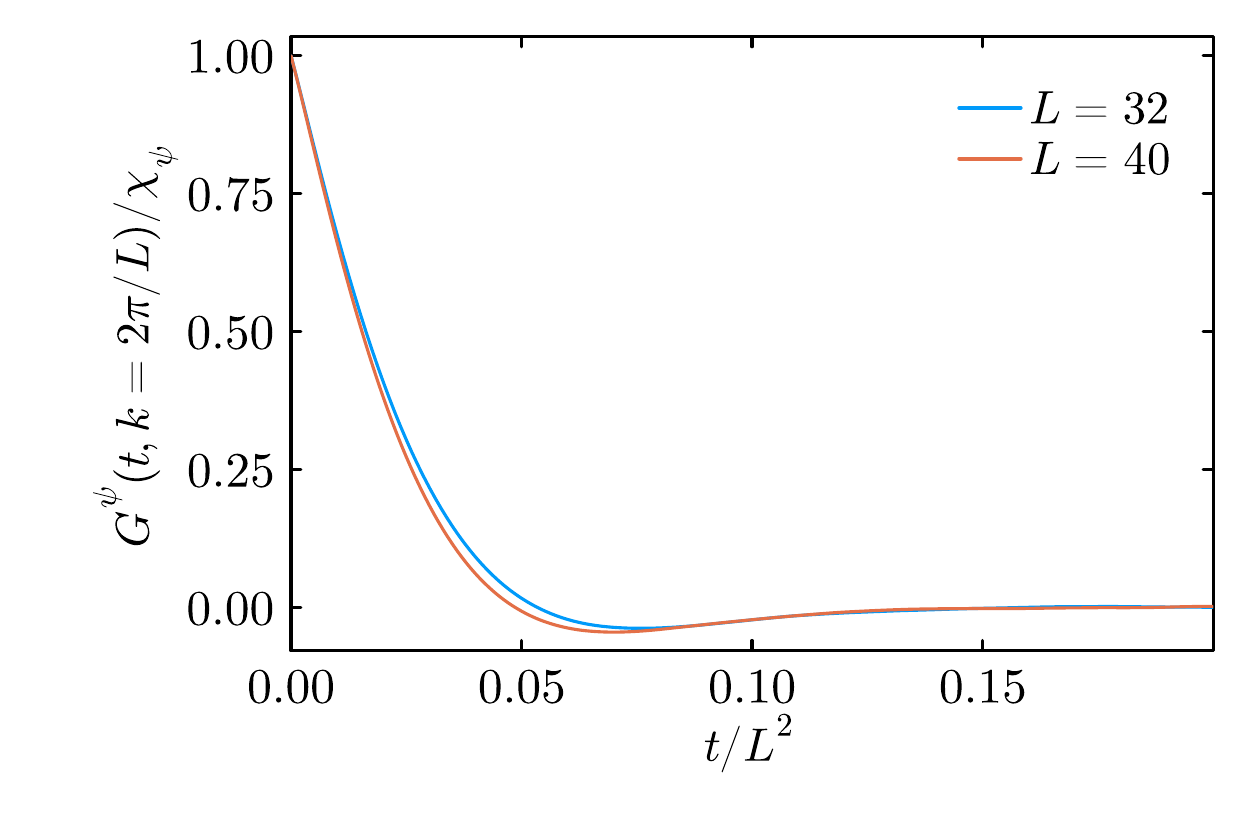}
\includegraphics[width=0.49\linewidth]{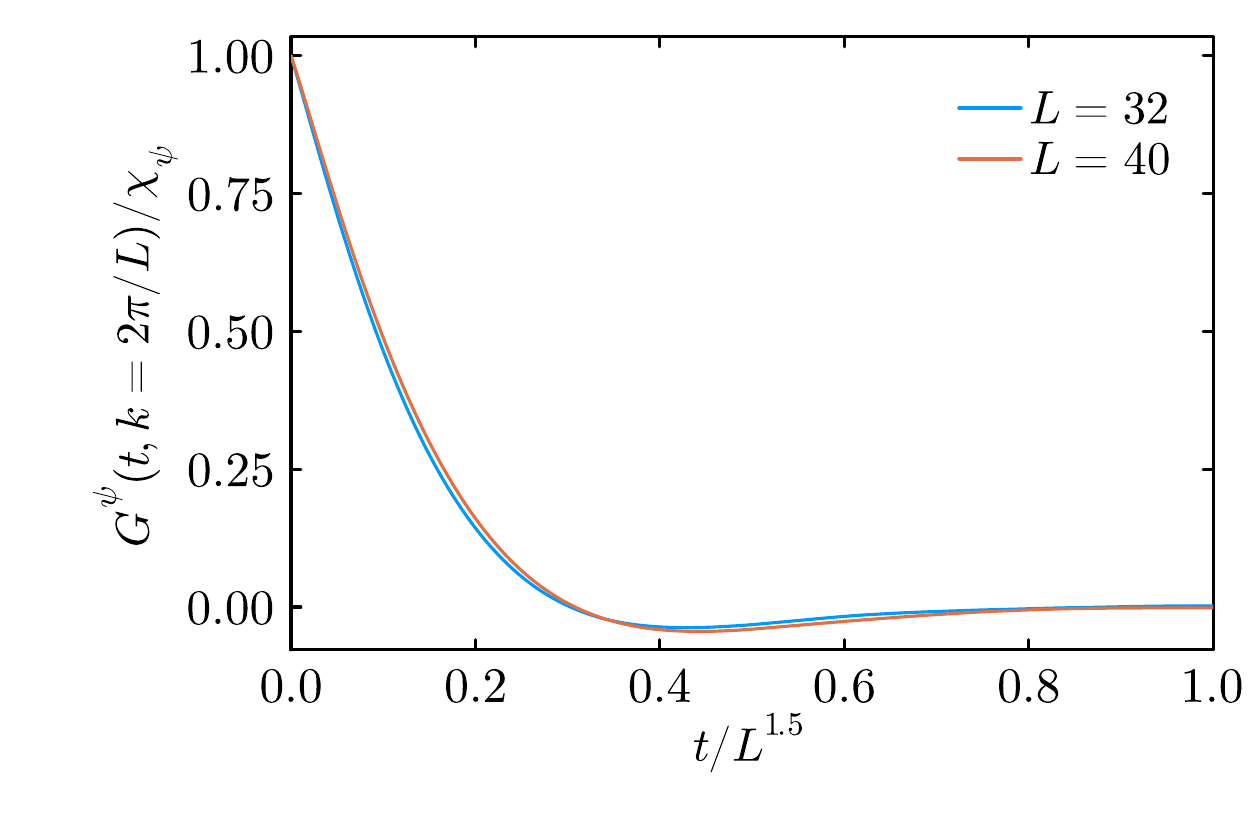}
\caption{Conserved density correlation function $G^\psi(t,k = 2\pi /L)$
at the critical point $m^2=m_c^2$ and $H=0$ for two different values 
of the lattice size, $L=32$ and $L=40$. The correlation function
is plotted as a function of the scaling variable $t/L^z$ for two
different values of the dynamic exponent, $z_\psi=2$ (left panel) and
$z_\psi=3/2$ (right panel). The best fit value of $z_\psi$, based on 
the scaling behavior at $G^\psi(t)/\chi_\psi = 0.3$, is $z_\psi = 
1.715 \pm 0.026$.}
\label{fig:Z_extr_psi}
\end{figure}
%%%%%%%%%%%%%%%%%%%%%%%%%%%%%%%%%%%%%%%%%%%%%%%%%%%%%%%%%%%%%%%%%%%%%%%%%%%%%%

 We study dynamical correlation functions of the order parameter 
field $\phi_a$ and the conserved density $\psi$. We work with 
spatial Fourier components 
\begin{align}
\phi_a(\vec{k})=\sum_{\vec{x}}\phi_a(\vec{x})\, 
  e^{i\vec{x}\cdot\vec{k}}\, , \quad
\psi(\vec{k})=\sum_{\vec{x}}\psi(\vec{x})\, 
  e^{i\vec{x}\cdot\vec{k}}\, ,
\end{align}
where on the lattice we have discrete wave numbers $\vec{k}=
2\pi \vec{m}/L$ with $m_i=0,\ldots,N-1$. We define the correlation
functions
\begin{align}
 G_{aa}^\phi(t,\vec{k}) = 
    \left\langle \phi_a(0,\vec{k})\phi_a(t,-\vec{k})\right\rangle  \, ,
    \quad
G^\psi(t,\vec{k}) = 
    \left\langle \psi(0,\vec{k})\psi(t,-\vec{k})\right\rangle  \, ,
\end{align}
where $G_{aa}^\phi$ contains no sum over $a$. We consider an external
field in the $a=1$ direction, $H_a=\delta_{a1}H$, and define the 
correlation functions in the Higgs and Goldstone boson direction, 
$G_\sigma(t,\vec{k})=G_{11}^\phi(t,\vec{k})$ and $G_\pi(t,\vec{k})
=G_{22}^\phi(t,\vec{k})$. Note that in the broken phase we always 
work at non-zero field $H$. For $H=0$ in the symmetric phase 
$O(2)$ symmetry implies $G_\sigma(t,\vec{k})=G_\pi(t,\vec{k})$.

%%%%%%%%%%%%%%%%%%%%%%%%%%%%%%%%%%%%%%%%%%%%%%%%%%%%%%%%%%%%%%%%%%%%%%%%%%%%%%
\begin{figure}[t]
\centering
\includegraphics[width=0.49\linewidth]{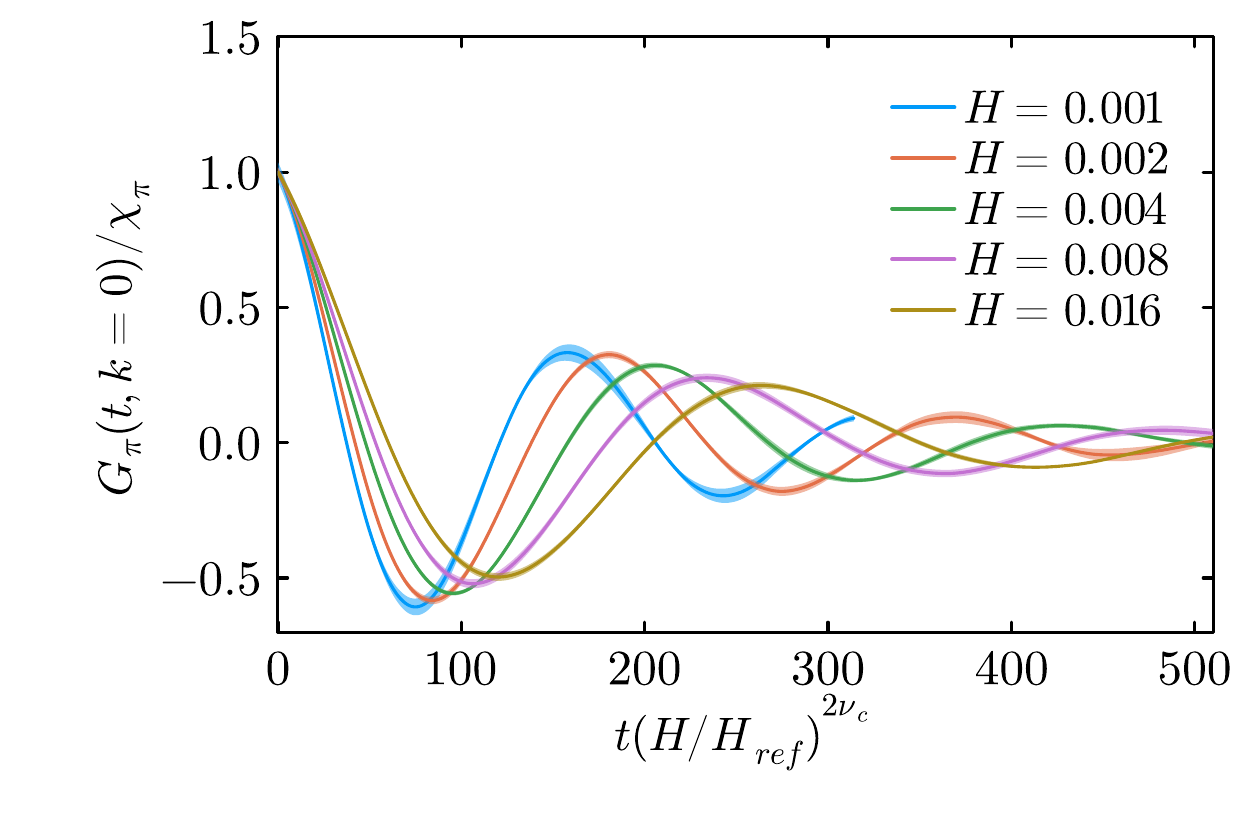}
\includegraphics[width=0.49\linewidth]{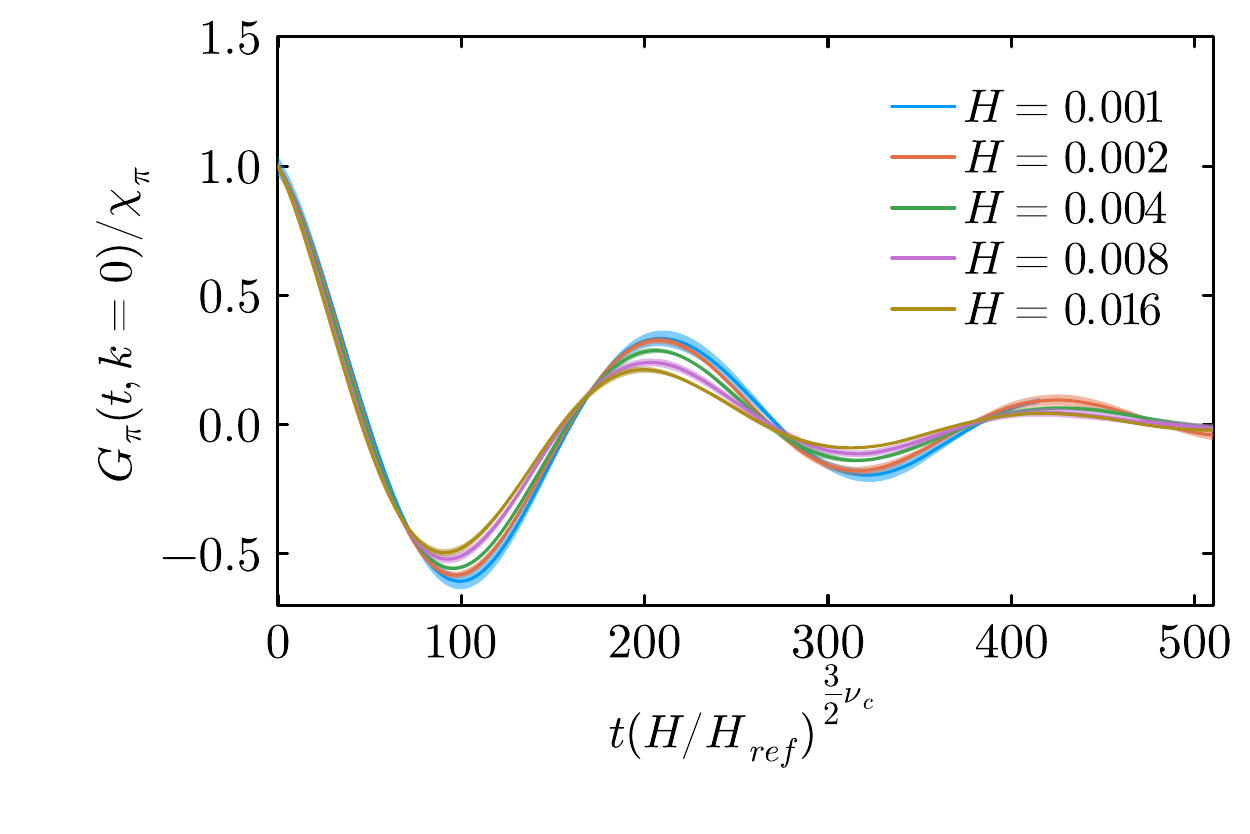}
\caption{Goldstone boson correlation function $G_\pi(t,0)$ at the
critical temperature $m^2=m_c^2$ for different values of the external
field $H$. We plot the correlation function as a function of the
scaling variable $th^{z\nu_c}$ for two different values of the 
dynamic exponent, $z=2$ (left panel) and $z=3/2$ (right panel). 
Here, $h=H/H_{\it ref}$ with $H_{ref} = 0.004$ and the lattice size 
is $L=40$. }
\label{fig:Z_x-check}
\end{figure}
%%%%%%%%%%%%%%%%%%%%%%%%%%%%%%%%%%%%%%%%%%%%%%%%%%%%%%%%%%%%%%%%%%%%%%%%%%%%%%

 Fig.~\ref{fig:Z_extr_phi} shows $G_\sigma(t,0)$ at the critical 
point $m^2=m_c^2$, $H=0$, for two different values of the lattice size,
$L=32$ and $L=40$. At the critical point the correlation length 
$\xi$ is only limited by $L$, and dynamical scaling implies that 
$G_\sigma(t,\vec{k};L)= \bar{G}_\sigma(t/L^z,\vec{k}L)$. The left
and right panel show two different scaling plots with $z=2$ and $z=3/2$,
and we observe that the value $z=3/2$ is clearly preferred. To be 
more quantitative we determined $z$ from the value of $t$ at which
$G_\sigma(t,0)=0.3$. We find $z = 1.51 \pm 0.14$ \footnote{
The value of the correlation function at which determine the scaling
parameter is a rough compromise between resolution and statistics. 
The stated error in $z$ is due to the statistical error on the 
correlation function. Note that $G_\psi(t,\vec{k}_1)$ falls of more 
quickly than $G_\sigma(t,0)$. As a result, the statistical error of 
$z$ determined from $G_\psi(t,\vec{k}_1)=0.3$ is smaller, but the 
sensitivity to genuine IR behavior is worse.}, 
which is in very good agreement with the prediction $z=3/2$ discussed 
in Sect.~\ref{sec:ModelF-dyn}. In Fig.~\ref{fig:Z_extr_psi} we show a 
similar scaling plot for the correlation function of $\psi$. In this 
case, the $\vec{k}=0$ mode reflects the initial condition for the 
integral of $\psi$, and carries no dynamical information. We show the 
correlator for the lowest momentum mode $\vec{k}_1$. Here, the best 
fit value shows some deviation from $z=3/2$. We find $z_\psi = 1.715
\pm 0.026$, where the quoted error is purely statistical.

%%%%%%%%%%%%%%%%%%%%%%%%%%%%%%%%%%%%%%%%%%%%%%%%%%%%%%%%%%%%%%%%%%%%%%%%%%%%%%
\begin{figure}[t]
\centering
\includegraphics[width=0.49\linewidth]{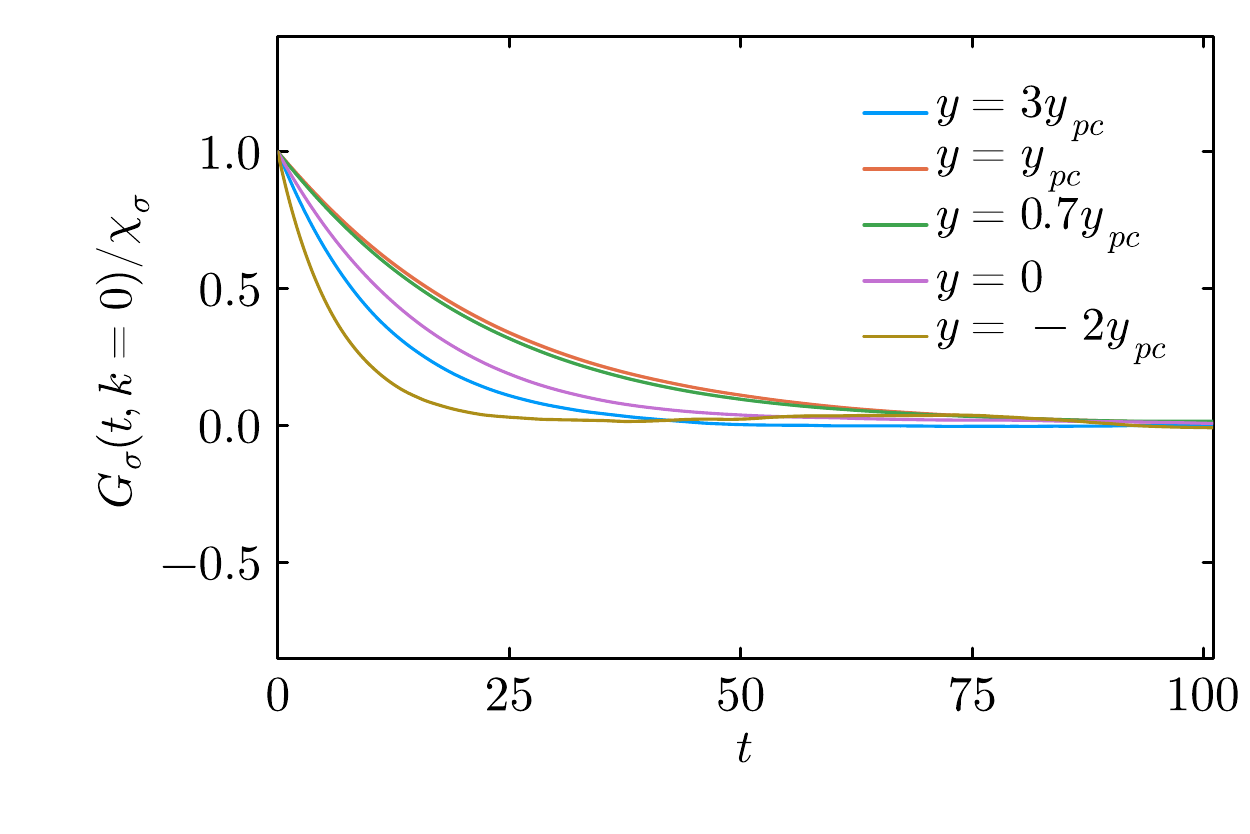}
\includegraphics[width=0.49\linewidth]{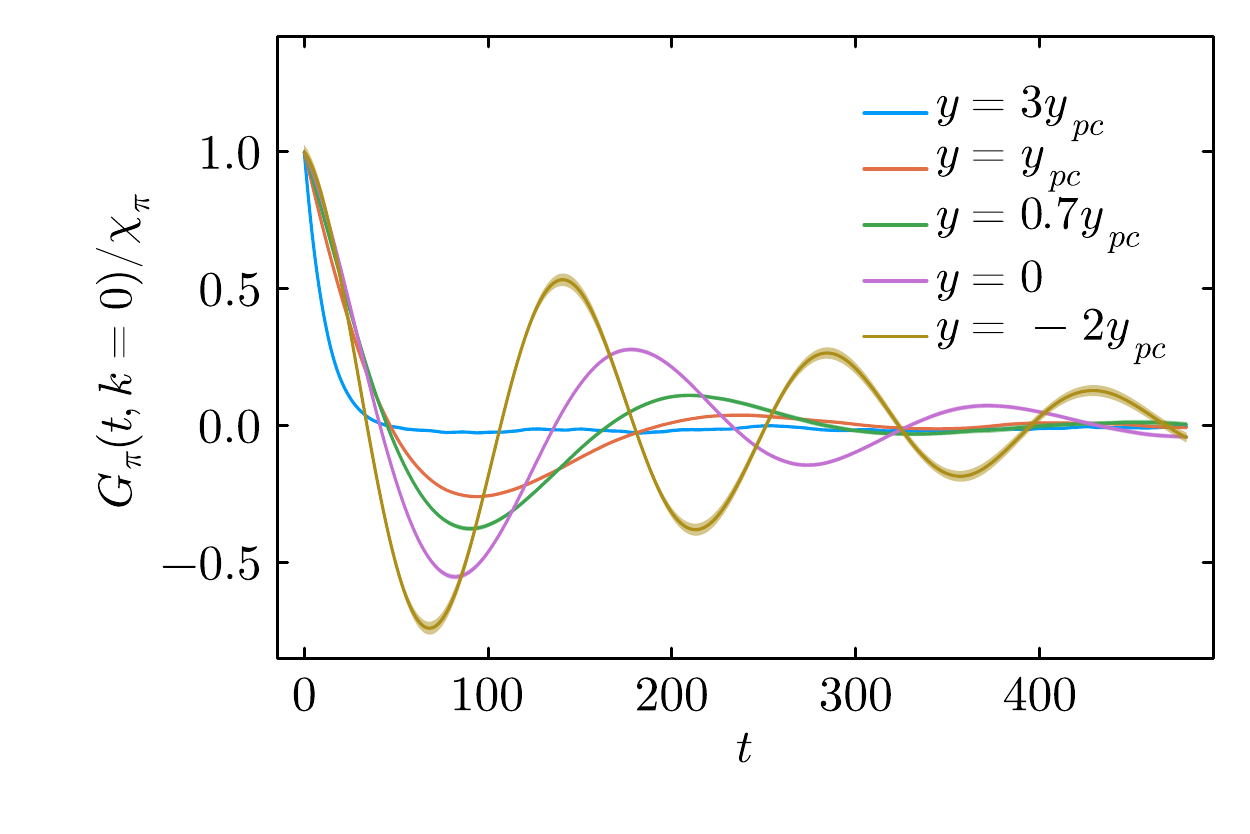}
\caption{Order parameter correlation functions $G_\sigma(t,0)$
(left panel) and $G_\pi(t,0)$ (right panel) for different values
of the universal variable $y$, see Fig.~\ref{fig:fgfchi}
and Table \ref{tab:compset}. The correlation functions are normalized 
to the respective susceptibilities.
\label{fig:phi2phi2}}
\end{figure}
%%%%%%%%%%%%%%%%%%%%%%%%%%%%%%%%%%%%%%%%%%%%%%%%%%%%%%%%%%%%%%%%%%%%%%%%%%%%%%

 In Fig.~\ref{fig:Z_x-check} we study the scaling with the external 
field at the critical temperature $m^2=m_c^2$. We show $G_\pi(t,0)$
as a function of the scaling variable $th^{z\nu_c}$, again comparing
two different values of the dynamic exponent, $z=2$ and $z=3/2$. Here,
we have taken the value of the static exponent $\xi\sim h^{-\nu_c}$
with $\nu_c=\nu/(\beta\delta)$ from Eq.~(\ref{crit-exp}). Again, we 
observe that the value $z=3/2$ is clearly preferred. 

\begin{figure}[t]
\centering
\includegraphics[width=0.5\linewidth]{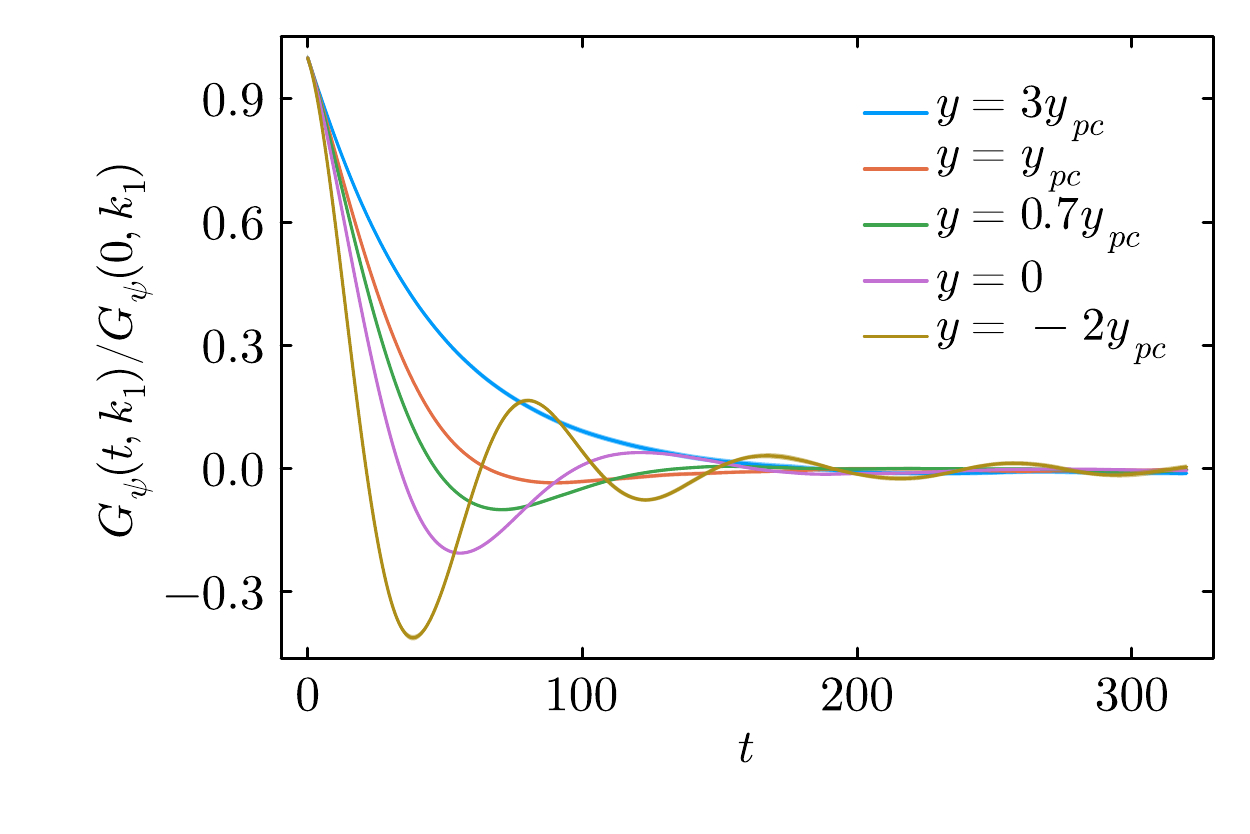}
\caption{Correlation function of the conserved density $G_\psi(t,\vec{k}_1)$
as a function of time for the lowest momentum mode $\vec{k}_1$. 
We show the result for several values of the universal
quantity $y$ as in the previous figure.}
\label{fig:psipsi}
\end{figure}

 In Figs.~\ref{fig:phi2phi2} and \ref{fig:psipsi} we show 
the correlation functions of $\phi_a$ and $\psi$ for different values 
of the universal variable $\bar{z}$ that enters the magnetic equation 
of state, see Fig.~\ref{fig:fgfchi}. Fig.~\ref{fig:phi2phi2} shows
the Higgs and Goldstone mode in the left and right panel, respectively. 
We observe that the Higgs mode is diffusive for all values of $y$,
but a propagating Goldstone mode emerges for $y < y_{pc}$. This 
mode becomes less damped as we move deeper into the broken phase. In
Fig.~\ref{fig:psipsi} we show the analogous result for the correlation
function of the conserved density. As in Fig.~\ref{fig:Z_extr_psi}
we show the correlator for the lowest momentum 
mode $\vec{k}_1$ Again, we observe that the correlation function
changes from diffusive behavior to exhibiting a propagating mode
as we move into the broken phase. Physically, this corresponds to 
the emergence of second sound at the superfluid phase transition.

%%%%%%%%%%%%%%%%%%%%%%%%%%%%%%%%%%%%%%%%%%%%%%%%%%%%%%%%%%%%%%%%%%%%%%%%%%%%%%
\subsection{Transport and Kubo relation}
\label{sec:Kubo}
%%%%%%%%%%%%%%%%%%%%%%%%%%%%%%%%%%%%%%%%%%%%%%%%%%%%%%%%%%%%%%%%%%%%%%%%%%%%%%

%%%%%%%%%%%%%%%%%%%%%%%%%%%%%%%%%%%%%%%%%%%%%%%%%%%%%%%%%%%%%%%%%%%%%%%%%%%%%%
%%%%%%%%%%%%%%%%%%%%%%%%%%%%%%%%%%%%%%%%%%%%%%%%%%%%%%%%%%%%%%%%%%%%%%%%%%%%%%

  The divergence of the thermal conductivity and the second
sound attenuation constant is an interesting and experimentally 
accessible manifestation of dynamic scaling near the superfluid
phase transition. There are (at least) two possible strategies
to access these transport coefficients. The first is to study
the asymptotic behavior of $G_\psi(t,\vec{k})$. In the symmetric
phase, we expect $G_\psi(t,\vec{k})\sim \exp(-tD_sk^2)$, and 
in the broken phase $G_\psi(t,\vec{k})\sim \sin(c_s kt+\varphi_s)
\exp(-D_stk^2)$. Here, $D_s=\kappa/C$ is the thermal diffusivity,
$c_s$ is the speed of sound, and $\varphi_s$ is a phase. For 
$\gamma_0=0$ we have $C=C_0$, but for $\gamma_0\neq 0$ we have 
to measure the thermal conductivity in order to extract $\kappa$
from $D_s$. As noted above, in the $O(2)$ model the specific heat
does not diverge, and the critical scaling of $D_s$ and $\kappa$
is the same. The difficulty with this strategy is that we have 
to identify a suitable window in which the correlation function (or its
envelope) decays exponentially, and in which the decay constant
scales as $k^2$. 

 A second strategy is based on the Kubo relation, which expresses
the thermal conductivity in terms of the time integrated current-current
correlation function \cite{Kubo:1957,Luttinger:1964zz,
Chattopadhyay:2025uqo},
\begin{align}
    \kappa = \int_0^\infty dt\, C_{\jmath\jmath}(t)\, , \quad
        C_{\jmath\jmath}(t) = \frac{1}{3TV} \, 
    \langle J_i(0) \, J^i(t) \rangle, 
\end{align}
where $J^i(t)=\int d^3x\, \jmath(t,\vec{x})$ and the current $\jmath$
is defined in Eq.~(\ref{psi-cons}). The stochastic current $\vec{\xi}$
contributes a delta-function that corresponds to the tree
level contribution $\kappa=\kappa_0$. The remainder of the integral
can be interpreted as the fluctuation induced thermal conductivity 
$\Delta\kappa$. A possible difficulty with the Kubo relation is 
that $C_{\jmath\jmath}$ may contain additional UV sensitive terms,
contributions of the form $C_{\jmath\jmath}(t)\sim t^{-\rho}$ with 
$\rho\geq 1$, which are sensitive to the lattice regulator. Terms
of this form are known to be present in the Kubo intgrand for the 
shear viscosity in a normal fluid \cite{Kovtun:2011np}, but they 
do not affect the extraction of critical exponents
\cite{Chattopadhyay:2025uqo}.

%%%%%%%%%%%%%%%%%%%%%%%%%%%%%%%%%%%%%%%%%%%%%%%%%%%%%%%%%%%%%%%%%%%%%%%%%%%%%%
\begin{figure}[t]
\centering
\includegraphics[width=0.49\linewidth]{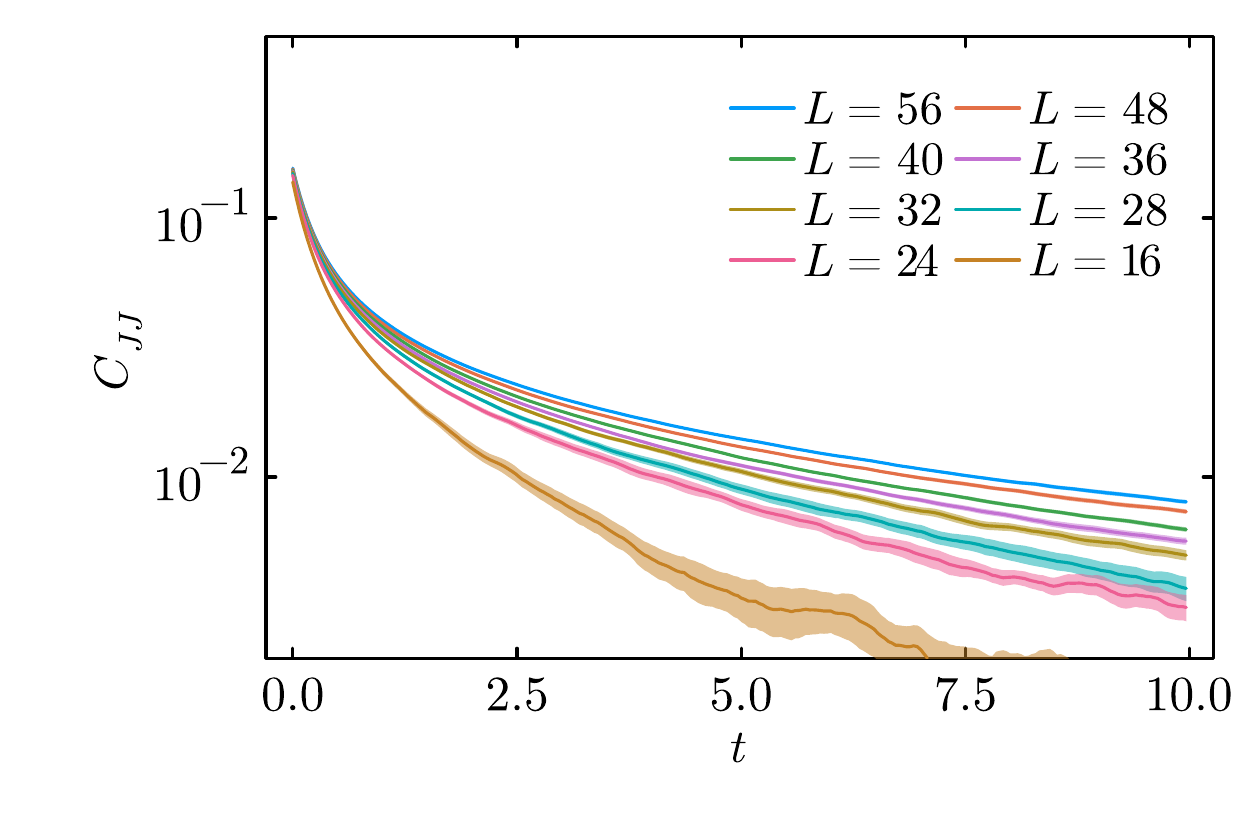}
\includegraphics[width=0.49\linewidth]{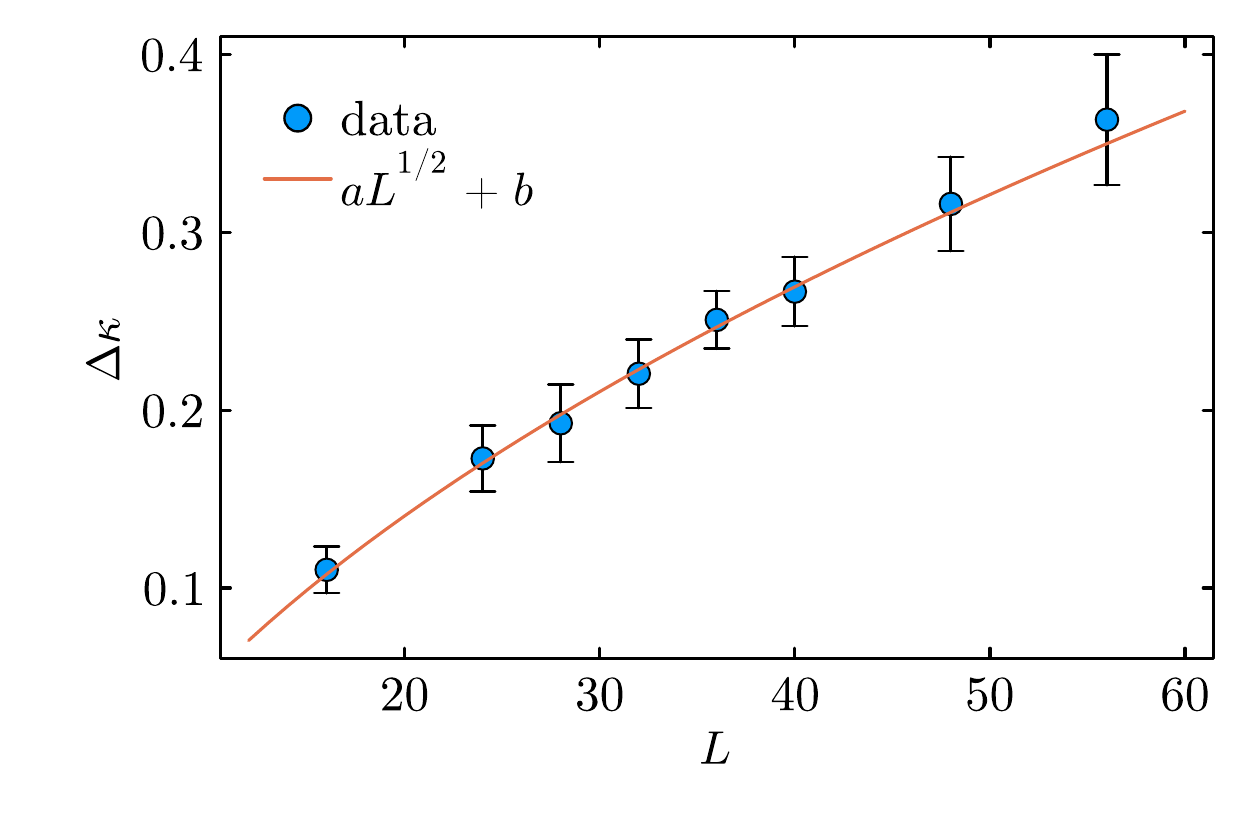}
\caption{Current-current correlation function $C_{\jmath\jmath}(t)$
(left panel) and fluctuation contribution to the thermal conductivity
(right panel). The correlation function as a function of time is 
shown for several values of the lattice size $L$. The Kubo integral
$\Delta\kappa$ is shown as a function of $L$, together with the 
fit $\Delta\kappa=aL^{1/2}+b$. The best fit parameters are 
$a=0.069\pm 0.009$ and $b = -0.17 \pm 0.046$.
}
\label{fig:JJ}
\end{figure}
%%%%%%%%%%%%%%%%%%%%%%%%%%%%%%%%%%%%%%%%%%%%%%%%%%%%%%%%%%%%%%%%%%%%%%%%%%%%%%

 We note that the dissipative current $\vec\jmath_{\it dis}=-\kappa
\vec{\nabla} (\delta {\cal H}/\delta \psi)$ vanishes when summed over 
all lattice sites. Therefore, we will consider the lattice integral 
of the current defined in Eq.~(\ref{cur-lat})
\begin{align}
\label{cur-lat-zm}
J_\mu(t) = g_0 \sum_{\vec{x}}\left( 
    \phi_1(t,\vec{x})\nabla^R_\mu\phi_2(t,\vec{x})
  - \phi_2(t,\vec{x})\nabla^R_\mu\phi_1(t,\vec{x})\right)\, .
\end{align}
Note that using the current $\jmath_\mu^L(t,\vec{x})$ leads to the 
same integrated quantity $J_\mu(t)$. In Fig.~\ref{fig:JJ} (left panel) 
we show the correlation function $C_{\jmath\jmath}(t)$ of the current 
in Eq.~(\ref{cur-lat-zm}). The calculation is performed at the critical 
point for a number of lattice sizes $L$. We observe that the short-time 
behavior is indeed insensitive to $L$, but that the long-time tail 
increases with $L$. In the right panel we show the fluctuation induced
contribution to the thermal conductivity $\Delta\kappa$. Because of the 
limited range in $L$ that we can access, the data do not determine the 
exponent $x_\kappa$ in the scaling relation $\Delta\kappa\sim L^{x_\kappa}$
with very high accuracy. The results are consistent with the prediction
$x_\kappa=1/2$ based on dynamic scaling and $z=3/2$. In particular, as 
shown in the right panel of Fig.~\ref{fig:JJ}, the data is very well 
described by $\Delta\kappa=aL^{1/2}+b$ with $a=0.069\pm 0.009$ and 
$b = -0.17 \pm 0.046$.

%%%%%%%%%%%%%%%%%%%%%%%%%%%%%%%%%%%%%%%%%%%%%%%%%%%%%%%%%%%%%%%%%%%%%%%%%%%%%%
\section{Conclusion and Outlook}
\label{sec:sum}
%%%%%%%%%%%%%%%%%%%%%%%%%%%%%%%%%%%%%%%%%%%%%%%%%%%%%%%%%%%%%%%%%%%%%%%%%%%%%%

  In this work, we have studied the critical dynamics of the superfluid
transition using numerical simulations of the model E truncation of 
model F. This model describes the coupled dynamics of a relaxational
order parameter mode and a diffusive heat current. We observe 
dynamical scaling with a critical exponent consistent with $z=3/2$, 
and the emergence of a second sound mode. A determination of the 
dynamic critical exponent based on finite size scaling of the order 
parameter correlation function gives $z=1.51\pm 0.14$. Extractions
of $z$ based on other correlation functions, or on the scaling with 
the external field, are consistent within error bars. The diffusivity 
of the second sound mode diverges as $D_s\sim \xi^{x_\kappa}$, where 
$x_\kappa$ is compatible with the scaling prediction $x_\kappa=1/2$.

 There are a number of compelling directions for future work. First, 
it would be interesting to test the truncations employed in our 
study. This includes the role of the coupling $\gamma_0$ and the 
imaginary part of the order parameter relaxation rate, $\Gamma_2$.
Going beyond this, one can investigate the role of fluctuations 
of the momentum density, which amounts to including shear modes
and ordinary, first sound, modes. 

  Second, there is an opportunity for quantitative comparison 
with experimental data on the dynamics of ultracold atomic gases
near the superfluid phase transition. As discussed in the introduction,
an enhancement of the second sound diffusivity has been observed
in an experiment that combines local thermometry with linear 
response \cite{Yan:2024}. An attempt to understand this behavior 
quantitatively requires mapping the parameters of the free 
energy functional Eq.~(\ref{Hamiltonian}) onto the equation of 
state, and determining the dynamical parameters $\Gamma_1$, $\Gamma_2$,
$\kappa$, and $g_0$. Numerical studies of this type have been
performed in the case of liquid helium \cite{Dohm:1987}, but 
the unitary Fermi gas has the advantage that the microscopic
interaction is simpler. The unitary gas also offers the 
opportunity to study out-of-equilibrium transitions such as 
quenches or transitions in an expanding gas, and to monitor 
fluctuations of the density and temperature with atomic 
resolution. 

\acknowledgments 
We are grateful to Eduardo Grossi and Derek Teaney for stimulating
discussions. This work is supported by the U.S. Department of Energy, 
Office of Science, Office of Nuclear Physics through Contract No. 
DE-SC0020081, DE-FG02-03ER41260, and DE22-SC0024622.  C.C. is supported 
by the Department of Space (DOS), Government of India.  
We acknowledge the computing resources provided by North Carolina State 
University High Performance Computing Services Core Facility 
(RRID:SCR-022168) and the computational resources provided by NCShare, 
which is supported by National Science Foundation (NSF) grants OAC-2201525, 
OAC-2201105, and OAC-2430141. We also thank Uthpala Herath for the 
computational support at NCShare.

\appendix
%%%%%%%%%%%%%%%%%%%%%%%%%%%%%%%%%%%%%%%%%%%%%%%%%%%%%%%%%%%%%%%%%%%%%%%%%%%%%%
\section{Metric factors and universal scaling}
\label{sec:App-H0}
\subsection{$H_0$ and $L_0$} 
%%%%%%%%%%%%%%%%%%%%%%%%%%%%%%%%%%%%%%%%%%%%%%%%%%%%%%%%%%%%%%%%%%%%%%%%%%%%%%

%%%%%%%%%%%%%%%%%%%%%%%%%%%%%%%%%%%%%%%%%%%%%%%%%%%%%%%%%%%%%%%%%%%%%%%%%%%%%%
\begin{figure}[t]
\centering
\includegraphics[width=0.45\linewidth]{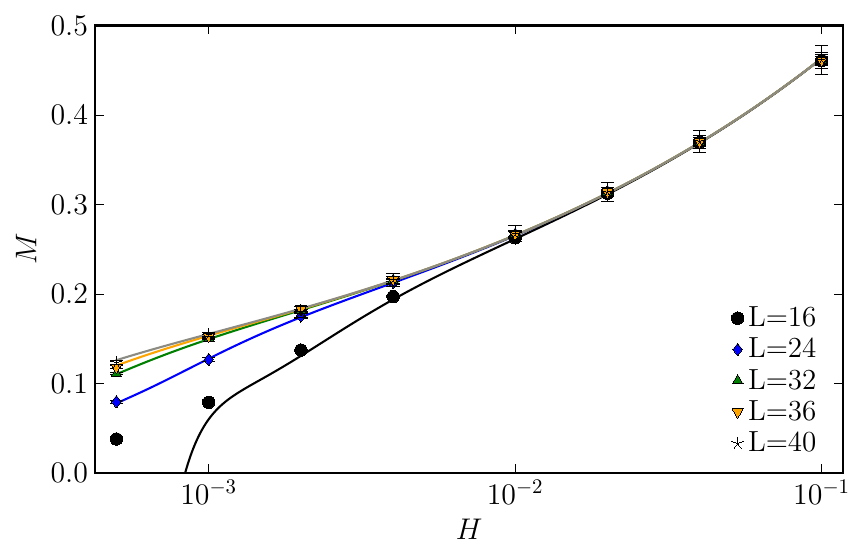}
\includegraphics[width=0.43\linewidth]{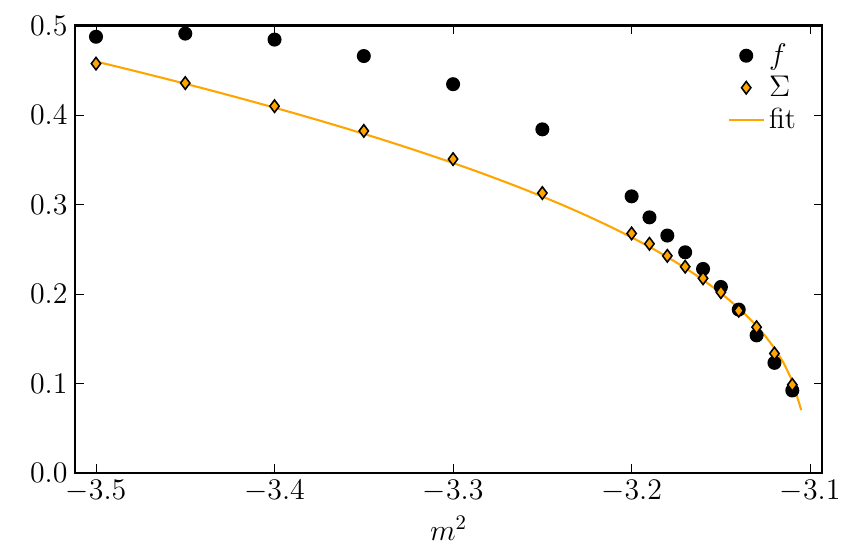}
 \caption{ Left panel: 
The simulation data (symbols) at critical temperature for various $H$ and 
$L$ and its fits (curves) by Eq.~\eqref{eq:MatTc}. Only $L>24$ are used 
to fit the parameters. 
Right panel: Goldstone boson decay constant, $f$ and order parameter, $\Sigma$ 
as a function of temperature, $m^2$. The fit is based on Eq.~\eqref{eq:sigma_t}.} 
\label{fig:metricfactors1}
\end{figure}
%%%%%%%%%%%%%%%%%%%%%%%%%%%%%%%%%%%%%%%%%%%%%%%%%%%%%%%%%%%%%%%%%%%%%%%%%%%%%%

The metric factors $H_0$ and $L_0$ are determined by coupling an external
field $H\equiv H_1$ along the $\phi_1$ direction. We measure the $H$ and $L$ 
dependence of the order parameter $M\equiv M_1$ defined in Eq.~(\ref{M-def}),
and fit the result to the universal magnetic equation of state
\begin{align}
\big\langle M \big\rangle  = \left(\frac{H}{H_0}\right)^{1/\delta}
    \left[
    f_G\!\left(y = 0,\; y_L = \frac{L_0}{L}
    \left(\frac{H}{H_0}\right)^{-\nu_c}\right)
    + C_H \left(\frac{H}{H_0}\right)^{\omega\nu_c}
    \right],
\label{eq:MatTc}
\end{align}
where $\nu_c \equiv \nu/(\beta\delta)$ and we have approximated $f_G^{(1)}
(0,y_L)\simeq C_H$. This fit determines the parameters $H_0$, $L_0$, 
and $C_H$. The function $f_G(y,y_L)$ is taken from a parametrization 
given in Ref.~\cite{Karsch:2023pga}. The critical exponents for the 
three-dimensional $O(2)$ universality class are collected in
Eq.~(\ref{crit-exp}).

 The fit is performed on lattices of size $L = 32, 36, 40$. To assess the 
quality of the fit, the scaling curves are compared against data at $L = 16$
and $L = 24$; see the left panel of Fig.~\ref{fig:metricfactors1}. The comparison
shows that only the smallest volume, $L = 16$, departs from the expected scaling
behavior at $H \lesssim 10^{-3}$, indicating the onset of finite-size corrections
beyond the scaling window. The best-fit values of the metric factors are
\begin{align}
%H0 -> 7.97325, L0 -> 0.566766, CH -> 0.642232}
%{0.165342, 0.00798737, 0.0365077}
    H_0 &= 7.97 \pm 0.17, \notag \\
    L_0 &= 0.567 \pm 0.008, \label{eq:metricfactors}\\
    C_H &= 0.64 \pm 0.04\,. \notag
\end{align}

%%%%%%%%%%%%%%%%%%%%%%%%%%%%%%%%%%%%%%%%%%%%%%%%%%%%%%%%%%%%%%%%%%%%%%%%%%%%%%%
\begin{figure}
\centering
\includegraphics[width=0.5\linewidth]{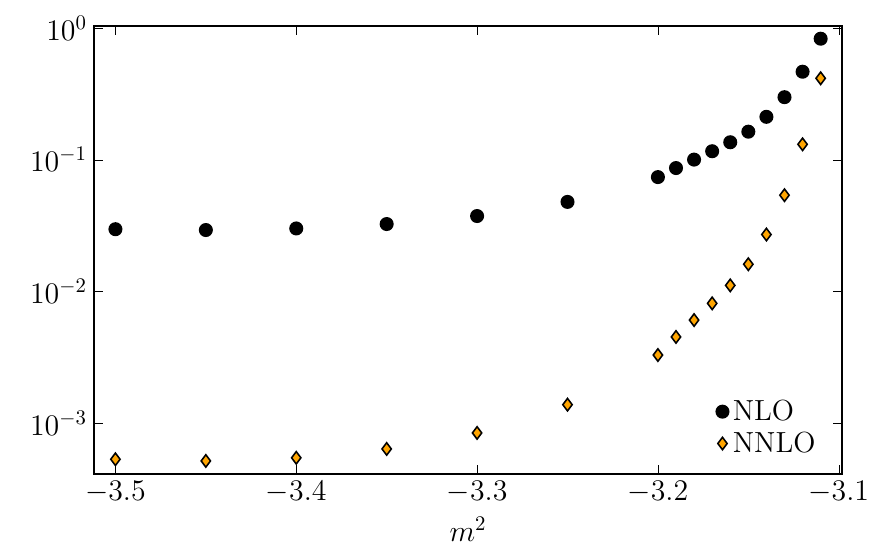}
\caption{An estimate of the series convergence in Eq.~\eqref{eq:M2_final}. 
LO term is $10^0$. The figure shows that the series expansion breaks down 
near the critical temperature and is no longer reliable. Thus to perform 
the fit in Right panel of  Fig.~\ref{fig:metricfactors1} we restricted the
range to $m^2\leq -3.15$, where NLO/LO $\lessapprox$ NNLO/NLO $\lessapprox 
10^{-1}$.  }
\label{fig:convtest}
\end{figure}
%%%%%%%%%%%%%%%%%%%%%%%%%%%%%%%%%%%%%%%%%%%%%%%%%%%%%%%%%%%%%%%%%%%%%%%%%%%%%%%

%%%%%%%%%%%%%%%%%%%%%%%%%%%%%%%%%%%%%%%%%%%%%%%%%%%%%%%%%%%%%%%%%%%%%%%%%%%%%%%
\subsection{Metric factor $\tau_0$}
%%%%%%%%%%%%%%%%%%%%%%%%%%%%%%%%%%%%%%%%%%%%%%%%%%%%%%%%%%%%%%%%%%%%%%%%%%%%%%%%

 The normalization factor $\tau_0$ can be extracted from the $m^2$ dependence
of the order parameter at zero field, see the right panel of 
Fig.~\ref{fig:metricfactors1}. The difficulty is that in order to use the 
definition in Eq.~(\ref{M-def}) we have to extrapolate the lattice results 
to infinite volume and zero field. An alternative  method that can be applied
directly at zero field is based on the following observation: In the broken 
phase at $H=0$, the finite-volume deviation of $\langle M^2 \rangle$ from 
its infinite-volume limit $\Sigma^2$ is driven by infrared fluctuations of 
the massless Goldstone modes. These contributions are systematically captured 
by the Euclidean effective theory in Eq.~(\ref{GB-EFT}), which organizes 
finite-size corrections as an expansion in powers of $(f^2 L)^{-1}$. In the 
regime $f^2(m^2)L \gg 1$, the connection between $\langle M^2 \rangle$ 
and $\Sigma^2$ takes the form \cite{Hasenfratz:1989pk}
\begin{align}
    \langle M^2 \rangle = \Sigma^2 \left(
    \rho_1^2 + \frac{2N\rho_2}{f^4 L^2} \right)
    + \mathcal{O}\!\left((f^2 L)^{-3}\right)\,.
\label{eq:effGBTheory}
\end{align}
The coefficients $\rho_1$ and $\rho_2$, which determine the $1/L$ corrections, 
are expressed through the geometry-dependent shape coefficients $\beta_1$ and 
$\beta_2$ as
\begin{align}
\rho_1 &= 1 + \frac{(N-1)\beta_1}{2f^2 L}
               - \frac{(N-1)(N-3)}{8f^4 L^2}\!\left(\beta_1^2 - 2\beta_2\right),
\label{rho1} \\
\rho_2 &= \frac{(N-1)\beta_2}{4}\,.
\label{rho2}
\end{align}
These shape coefficients arise from discrete Fourier sums over the momentum modes 
of a cubic box of side length $L$ and evaluate numerically to
\begin{align}
    \beta_1 = 0.225785\,, \qquad \beta_2 = 0.010608\,.
    \label{eq:betas}
\end{align}
Setting $N = 2$ and substituting these values, the expansion simplifies to
\begin{align}
\langle M^2 \rangle = \Sigma^2
 \left[
     1 + \underbrace{\frac{0.225785}{f^2 L}}_{\mathrm{NLO}}
       + \underbrace{\frac{0.0307934}{f^4 L^2}}_{\mathrm{NNLO}}
       + \mathcal{O}\!\left((f^2 L)^{-3}\right)
 \right],
\label{eq:M2_final}
\end{align}
which provides the basis for extracting $\Sigma$ from finite-volume lattice data.

 As $m^2 \to m^2_c$, the Goldstone boson decay constant vanishes, $f \to 0$, and 
the expansion in Eq.~\eqref{eq:M2_final} breaks down. It is therefore necessary 
to restrict the analysis to values of $m^2$ for which the series remains 
convergent. To delineate this region, we display the NLO and NNLO contributions 
separately in Fig.~\ref{fig:convtest}. A comparison of the two orders indicates 
that the expansion is reliable for $m^2 < -3.15$, which we adopt as the boundary 
of the fitting range. Within this range, we fit $\Sigma(m^2)$ to the expected 
power-law behaviour near the critical point in the low-temperature phase,
\begin{align}
\Sigma = \left(-\frac{\tau}{\tau_0}\right)^{\beta}
  \left[1 + C_T \left(-\frac{\tau}{\tau_0}\right)^{\omega\nu}\right],
\qquad \tau \equiv m^2 - m_c^2\,.
\label{eq:sigma_t}
\end{align}
The best-fit values of the free parameters are
\begin{align}
    \tau_0 &= 4.8\,, \notag \\
    C_T &= 0.37\,.
\end{align}
The scaling variable $z$ is thereby fully determined,
\begin{align}
    y = \frac{m^2 - m_c^2}{\tau_0}
        \left(\frac{H}{H_0}\right)^{- \frac{1} {\beta\delta} },
\label{eq:scalingvariable}
\end{align}
with metric factors $\tau_0 = 4.8$ and $H_0 = 7.97$.

\bibliography{bib.bib}

@article{Chester:2019ifh,
    author = "Chester, Shai M. and Landry, Walter and Liu, Junyu and Poland, David and Simmons-Duffin, David and Su, Ning and Vichi, Alessandro",
    title = "{Carving out OPE space and precise $O(2)$ model critical exponents}",
    eprint = "1912.03324",
    archivePrefix = "arXiv",
    primaryClass = "hep-th",
    reportNumber = "CALT-TH-2019-051",
    doi = "10.1007/JHEP06(2020)142",
    journal = "JHEP",
    volume = "06",
    pages = "142",
    year = "2020"
}

@article{Karsch:2023pga,
    author = "Karsch, Frithjof and Neumann, Marius and Sarkar, Mugdha",
    title = "{Scaling functions of the three-dimensional Z(2), O(2), and O(4) models and their finite-size dependence in an external field}",
    eprint = "2304.01710",
    archivePrefix = "arXiv",
    primaryClass = "hep-lat",
    doi = "10.1103/PhysRevD.108.014505",
    journal = "Phys. Rev. D",
    volume = "108",
    number = "1",
    pages = "014505",
    year = "2023"
}

@article{Schafer:2009dj,
    author = {Sch\"afer, Thomas and Teaney, Derek},
    title = "{Nearly Perfect Fluidity: From Cold Atomic Gases to Hot 
    Quark Gluon Plasmas}",
    eprint = "0904.3107",
    archivePrefix = "arXiv",
    primaryClass = "hep-ph",
    doi = "10.1088/0034-4885/72/12/126001",
    journal = "Rept. Prog. Phys.",
    volume = "72",
    pages = "126001",
    year = "2009"
}

@article{Florio:2021jlx,
    author = "Florio, Adrien and Grossi, Eduardo and Soloviev, Alexander and Teaney, Derek",
    title = "{Dynamics of the $O(4)$ critical point in QCD}",
    eprint = "2111.03640",
    archivePrefix = "arXiv",
    primaryClass = "hep-lat",
    journal = "Phys. Rev. D",
    volume = "105",
    number = "5",
    pages = "054512",
    year = "2022"
}

@article{Chattopadhyay:2024jlh,
    author = "Chattopadhyay, Chandrodoy and Ott, Josh 
    and Sch{\"a}fer, Thomas and Skokov, Vladimir V.",
    title = "{Simulations of Stochastic Fluid Dynamics near a 
    Critical Point in the Phase Diagram}",
    eprint = "2403.10608",
    archivePrefix = "arXiv",
    primaryClass = "nucl-th",
    doi = "10.1103/PhysRevLett.133.032301",
    journal = "Phys. Rev. Lett.",
    volume = "133",
    number = "3",
    pages = "032301",
    year = "2024"
}

@article{Hohenberg:1977ym,
    author = "Hohenberg, P. C. and Halperin, B. I.",
    title = "{Theory of Dynamic Critical Phenomena}",
    doi = "10.1103/RevModPhys.49.435",
    journal = "Rev. Mod. Phys.",
    volume = "49",
    pages = "435--479",
    year = "1977"
}

@article{Halperin:1976,
  title = {Renormalization-group treatment of the critical 
  dynamics of superfluid helium, the isotropic antiferromagnet, 
  and the easy-plane ferromagnet},
  author = {Halperin, B. I. and Hohenberg, P. C. and 
  Siggia, E. D.},
  journal = {Phys. Rev. B},
  volume = {13},
  issue = {3},
  pages = {1299--1328},
  numpages = {0},
  year = {1976},
  month = {Feb},
  publisher = {American Physical Society},
  doi = {10.1103/PhysRevB.13.1299},
  url = {https://link.aps.org/doi/10.1103/PhysRevB.13.1299}
}

@article{Bzdak:2019pkr,
    author = "Bzdak, Adam and Esumi, Shinichi and Koch, Volker and 
    Liao, Jinfeng and Stephanov, Mikhail and Xu, Nu",
    title = "{Mapping the Phases of Quantum Chromodynamics with 
    Beam Energy Scan}",
    eprint = "1906.00936",
    archivePrefix = "arXiv",
    primaryClass = "nucl-th",
    journal = "Phys. Rept.",
    volume = "853",
    pages = "1--87",
    year = "2020"
}

@article{Rajagopal:1992qz,
    author = "Rajagopal, Krishna and Wilczek, Frank",
    title = "{Static and dynamic critical phenomena at a second order QCD phase transition}",
    eprint = "hep-ph/9210253",
    archivePrefix = "arXiv",
    reportNumber = "PUPT-1347, IASSNS-HEP-92-60",
    doi = "10.1016/0550-3213(93)90502-G",
    journal = "Nucl. Phys. B",
    volume = "399",
    pages = "395--425",
    year = "1993"
}

@article{Zinn-Justin:2002ecy,
    author = "Zinn-Justin, Jean",
    title = "{Quantum field theory and critical phenomena}",
    journal = "Int. Ser. Monogr. Phys.",
    volume = "113",
    pages = "1--1054",
    year = "2002"
}

@article{Folk:2006ve,
    author = "Folk, R. and Moser, Hans-Guenther",
    title = "{Critical dynamics: a field-theoretical approach}",
    doi = "10.1088/0305-4470/39/24/R01",
    journal = "J. Phys. A",
    volume = "39",
    pages = "R207--R313",
    year = "2006"
}

@article{Schaefer:2022bfm,
    author = "Sch{\"a}fer, Thomas and Skokov, Vladimir",
    title = "{Dynamics of non-Gaussian fluctuations in model A}",
    eprint = "2204.02433",
    archivePrefix = "arXiv",
    primaryClass = "nucl-th",
    journal = "Phys. Rev. D",
    volume = "106",
    number = "1",
    pages = "014006",
    year = "2022"
}

@article{Chattopadhyay:2023jfm,
    author = "Chattopadhyay, Chandrodoy and Ott, Josh and
    Sch{\"a}fer, Thomas and Skokov, Vladimir",
    title = "{Dynamic scaling of order parameter fluctuations in 
    model B}",
    eprint = "2304.07279",
    archivePrefix = "arXiv",
    primaryClass = "nucl-th",
    journal = "Phys. Rev. D",
    volume = "108",
    number = "7",
    pages = "074004",
    year = "2023"
}

@article{Shu:1988,
  title={Efficient implementation of essentially non-oscillatory shock-
         capturing schemes},
  author={Shu, Chi-Wang and Osher, Stanley},
  journal={Journal of Computational Physics},
  volume={77},
  number={2},
  pages={439--471},
  year={1988},
  publisher={Elsevier}
}

@article{Kovtun:2011np,
    author = "Kovtun, Pavel and Moore, Guy D. and Romatschke, Paul",
    title = "{The stickiness of sound: An absolute lower limit on viscosity and the breakdown of second order relativistic hydrodynamics}",
    eprint = "1104.1586",
    archivePrefix = "arXiv",
    primaryClass = "hep-ph",
    journal = "Phys. Rev. D",
    volume = "84",
    pages = "025006",
    year = "2011"
}

@article{Adams:2012th,
    author = {Adams, Allan and Carr, Lincoln D. and Sch{\"a}fer, Thomas and Steinberg, Peter and Thomas, John E.},
    title = "{Strongly Correlated Quantum Fluids: Ultracold Quantum Gases, Quantum Chromodynamic Plasmas, and Holographic Duality}",
    eprint = "1205.5180",
    archivePrefix = "arXiv",
    primaryClass = "hep-th",
    doi = "10.1088/1367-2630/14/11/115009",
    journal = "New J. Phys.",
    volume = "14",
    pages = "115009",
    year = "2012"
}

@article{Schaefer:2014awa,
    author = "Sch{\"a}fer, Thomas",
    title = "{Fluid Dynamics and Viscosity in Strongly Correlated Fluids}",
    eprint = "1403.0653",
    archivePrefix = "arXiv",
    primaryClass = "hep-ph",
    doi = "10.1146/annurev-nucl-102313-025439",
    journal = "Ann. Rev. Nucl. Part. Sci.",
    volume = "64",
    pages = "125--148",
    year = "2014"
}

@article{Zwerger:2016,
    author = {{Zwerger}, W.},
    title = "{Strongly Interacting Fermi Gases}",
    eprint = {1608.00457},
    primaryClass = {cond-mat.quant-gas},
    note = {Proceedings of the International School of
            Physics “Enrico Fermi,” Course 191, edited by 
            M. Inguscio, W. Ketterle and S. Stringari, 
            IOS Press, Amsterdam (2014)},
    year = 2016
    }

@article{Ku:2012,
    title={Revealing the superfluid lambda transition in the universal 
    thermodynamics of a unitary Fermi gas},
    author={Ku, Mark JH and Sommer, Ariel T and Cheuk, Lawrence W and 
    Zwierlein, Martin W},
    journal={Science},
    volume={335},
    number={6068},
    pages={563--567},
    year={2012}
}

@article{Bulgac:2008zz,
    author = "Bulgac, Aurel and Drut, Joaquin E. and Magierski, Piotr",
    title = "{Quantum Monte Carlo simulations of the BCS-BEC crossover at finite 
    temperature}",
    eprint = "0803.3238",
    archivePrefix = "arXiv",
    primaryClass = "cond-mat.stat-mech",
    doi = "10.1103/PhysRevA.78.023625",
    journal = "Phys. Rev. A",
    volume = "78",
    pages = "023625",
    year = "2008"
}

@article{Yan:2024,
    author = {{Yan}, Zhenjie and {Patel}, Parth B. and {Mukherjee}, Biswaroop 
        and {Vale}, Chris J. and {Fletcher}, Richard J. and {Zwierlein}, 
        Martin W.},
    title = "{Thermography of the superfluid transition in a strongly 
        interacting Fermi gas}",
    journal = {Science},
    year = 2024,
    volume = {383},
    number = {6683},
    pages = {629-633},
    doi = {10.1126/science.adg3430},
    eprint = {2212.13752},
    primaryClass = {cond-mat.quant-gas}
}

@article{Dean:2003,
    title = {Pairing in nuclear systems: from neutron stars to finite nuclei},
    author = {Dean, D. J. and Hjorth-Jensen, M.},
    journal = {Rev. Mod. Phys.},
    volume = {75},
    issue = {2},
    pages = {607--656},
    year = {2003},
    doi = {10.1103/RevModPhys.75.607},
    url = {https://link.aps.org/doi/10.1103/RevModPhys.75.607}
}

@article{Alford:2007xm,
    author = {Alford, Mark G. and Schmitt, Andreas and Rajagopal, Krishna and 
    Sch{\"a}fer, Thomas},
    title = "{Color superconductivity in dense quark matter}",
    eprint = "0709.4635",
    archivePrefix = "arXiv",
    primaryClass = "hep-ph",
    doi = "10.1103/RevModPhys.80.1455",
    journal = "Rev. Mod. Phys.",
    volume = "80",
    pages = "1455--1515",
    year = "2008"
}

@article{Florio:2025zqv,
    author = "Florio, Adrien and Grossi, Eduardo and Mazeliauskas, Aleksas and Soloviev, Alexander and Teaney, Derek",
    title = "{Supercooled Goldstone Bosons at the QCD Chiral Phase Transition}",
    eprint = "2504.03516",
    archivePrefix = "arXiv",
    primaryClass = "hep-ph",
    doi = "10.1103/wyn4-ncdc",
    journal = "Phys. Rev. Lett.",
    volume = "135",
    number = "24",
    pages = "242303",
    year = "2025"
}

@article{Dohm:1979,
  title={Density correlation function and dynamic transient 
  exponents for liquid helium at and above T $\lambda$},
  author={Dohm, Volker},
  journal={Zeitschrift f{\"u}r Physik B Condensed Matter},
  volume={33},
  number={1},
  pages={79--95},
  year={1979},
  publisher={Springer}
}

@article{Dohm:1987,
  title={Renormalization-group theory of critical phenomena near the Lambda 
  transition of4He},
  author={Dohm, V},
  journal={Journal of low Temperature Physics},
  volume={69},
  number={1},
  pages={51--75},
  year={1987},
  publisher={Springer}
}

@article{Dohm:1991,
  title = {Renormalization-group flow equations of model F},
  author = {Dohm, Volker},
  journal = {Phys. Rev. B},
  volume = {44},
  issue = {6},
  pages = {2697--2712},
  numpages = {0},
  year = {1991},
  month = {Aug},
  publisher = {American Physical Society},
  doi = {10.1103/PhysRevB.44.2697},
  url = {https://link.aps.org/doi/10.1103/PhysRevB.44.2697}
}

@article{Dohm:2006,
  title = {Model $F$ in two-loop order and the thermal conductivity near the superfluid transition of $^{4}\mathrm{He}$},
  author = {Dohm, Volker},
  journal = {Phys. Rev. B},
  volume = {73},
  issue = {9},
  pages = {092503},
  numpages = {4},
  year = {2006},
  month = {Mar},
  publisher = {American Physical Society},
  doi = {10.1103/PhysRevB.73.092503},
  url = {https://link.aps.org/doi/10.1103/PhysRevB.73.092503}
}

@article{Haussmann:1999,
    title = {Liquid ${}^{4}\mathrm{He}$ near the superfluid transition in the presence of a heat current and gravity},
  author = {Haussmann, Rudolf},
  journal = {Phys. Rev. B},
  volume = {60},
  issue = {17},
  pages = {12349--12372},
  numpages = {0},
  year = {1999},
  month = {Nov},
  publisher = {American Physical Society},
  doi = {10.1103/PhysRevB.60.12349},
  url = {https://link.aps.org/doi/10.1103/PhysRevB.60.12349}
}

@article{Krech:1999,
  title = {Spin-dynamics simulations of the three-dimensional $\mathrm{XY}$ model: Structure factor and transport properties},
  author = {Krech, M. and Landau, D. P.},
  journal = {Phys. Rev. B},
  volume = {60},
  issue = {5},
  pages = {3375--3387},
  numpages = {0},
  year = {1999},
  month = {Aug},
  publisher = {American Physical Society},
  doi = {10.1103/PhysRevB.60.3375},
  url = {https://link.aps.org/doi/10.1103/PhysRevB.60.3375}
}

@article{Stoof:2001,
  title={Dynamics of fluctuating Bose--Einstein condensates},
  author={Stoof, HTC and Bijlsma, MJ},
  journal={Journal of low Temperature Physics},
  volume={124},
  number={3},
  pages={431--442},
  year={2001},
  publisher={Springer}
}

@article{Gardiner:2002,
    doi = {10.1088/0953-4075/35/6/310},
    url = {https://doi.org/10.1088/0953-4075/35/6/310},
    year = {2002},
    month = {mar},
    publisher = {},
    volume = {35},
    number = {6},
    pages = {1555},
    author = {C W Gardiner and J R Anglin and T I A Fudge},
    title = {The stochastic Gross-Pitaevskii equation},
    journal = {Journal of Physics B: Atomic, Molecular 
    and Optical Physics},
}

@article{Hasenfratz:1989pk,
    author = "Hasenfratz, P. and Leutwyler, H.",
    title = "{Goldstone Boson Related Finite Size Effects in 
    Field Theory and Critical Phenomena With O($N$) Symmetry}",
    reportNumber = "BUTP-89/28-BERN",
    doi = "10.1016/0550-3213(90)90603-B",
    journal = "Nucl. Phys. B",
    volume = "343",
    pages = "241--284",
    year = "1990"
}

@article{Fisher:1973,
  title = {Helicity Modulus, Superfluidity, and Scaling in 
  Isotropic Systems},
  author = {Fisher, Michael E. and Barber, Michael N. and Jasnow, 
  David},
  journal = {Phys. Rev. A},
  volume = {8},
  issue = {2},
  pages = {1111--1124},
  numpages = {0},
  year = {1973},
  month = {Aug},
  publisher = {American Physical Society},
  doi = {10.1103/PhysRevA.8.1111},
  url = {https://link.aps.org/doi/10.1103/PhysRevA.8.1111}
}

@article{Tauber:2014,
  title = {Perturbative Field-Theoretical Renormalization Group 
  Approach to Driven-Dissipative Bose-Einstein Criticality},
  author = {T\"auber, Uwe C. and Diehl, Sebastian},
  journal = {Phys. Rev. X},
  volume = {4},
  issue = {2},
  pages = {021010},
  numpages = {21},
  year = {2014},
  month = {Apr},
  publisher = {American Physical Society},
  doi = {10.1103/PhysRevX.4.021010},
  url = {https://link.aps.org/doi/10.1103/PhysRevX.4.021010}
}

@article{Kadanoff:1963,
    title = {Hydrodynamic equations and correlation functions},
    journal = {Annals of Physics},
    volume = {24},
    pages = {419-469},
    year = {1963},
    issn = {0003-4916},
    doi = {https://doi.org/10.1016/0003-4916(63)90078-2},
    author = {Leo P Kadanoff and Paul C Martin},
}

@article{Campostrini:2000iw,
    author = "Campostrini, Massimo and Hasenbusch, Martin and Pelissetto, 
    Andrea and Rossi, Paolo and Vicari, Ettore",
    title = "{Critical behavior of the three-dimensional xy universality 
    class}",
    eprint = "cond-mat/0010360",
    archivePrefix = "arXiv",
    reportNumber = "IFUP-TH-2000-31",
    doi = "10.1103/PhysRevB.63.214503",
    journal = "Phys. Rev. B",
    volume = "63",
    pages = "214503",
    year = "2001"
}

@article{Kubo:1957,
    author = {Kubo, Ryogo},
    title = "{Statistical-Mechanical Theory of Irreversible 
    Processes. I. General Theory and Simple Applications to 
    Magnetic and Conduction Problems}",
    journal = {Journal of the Physical Society of Japan},
    volume = {12},
    number = {6},
    pages = {570-586},
    year = {1957},
    doi = {10.1143/JPSJ.12.570},
}

@article{Luttinger:1964zz,
    author = "Luttinger, J. M.",
    title = "{Theory of Thermal Transport Coefficients}",
    doi = "10.1103/PhysRev.135.A1505",
    journal = "Phys. Rev.",
    volume = "135",
    pages = "A1505--A1514",
    year = "1964"
}

@article{Chattopadhyay:2025uqo,
    author = "Chattopadhyay, Chandrodoy and Ott, Josh and 
    Sch{\"a}fer, Thomas and Skokov, Vladimir V.",
    title = "{Transport properties of stochastic fluids}",
    eprint = "2510.12557",
    archivePrefix = "arXiv",
    primaryClass = "nucl-th",
    reportNumber = "MIT-CTP/5941",
    doi = "10.1103/kltq-qb4t",
    journal = "Phys. Rev. D",
    volume = "112",
    number = "11",
    pages = "114026",
    year = "2025"
}

@article{Bloch:2008,
   title={Many-body physics with ultracold gases},
   volume={80},
   ISSN={1539-0756},
   url={http://dx.doi.org/10.1103/RevModPhys.80.885},
   DOI={10.1103/revmodphys.80.885},
   number={3},
   journal={Reviews of Modern Physics},
   publisher={American Physical Society (APS)},
   author={Bloch, Immanuel and Dalibard, Jean and Zwerger, Wilhelm},
   year={2008},
   month=jul, 
   pages={885–964} 
}

@article{Page:2011,
  title = {Rapid Cooling of the Neutron Star in Cassiopeia A Triggered by 
  Neutron Superfluidity in Dense Matter},
  author = {Page, Dany and Prakash, Madappa and Lattimer, James M. and 
  Steiner, Andrew W.},
  journal = {Phys. Rev. Lett.},
  volume = {106},
  issue = {8},
  pages = {081101},
  numpages = {4},
  year = {2011},
  month = {Feb},
  publisher = {American Physical Society},
  doi = {10.1103/PhysRevLett.106.081101},
  url = {https://link.aps.org/doi/10.1103/PhysRevLett.106.081101}
}

\end{document}